\documentclass{aa}  

\usepackage{graphicx}
\usepackage{txfonts}
\usepackage{graphicx}	
\usepackage{amsmath}	
\usepackage{amssymb}	
\usepackage{siunitx}	%
\usepackage{enumitem}	%
\usepackage{subcaption}

\DeclareSIUnit[]\Mj
 {\text{\ensuremath{M_{\textup{J}}}}}
\DeclareSIUnit[]\Rj
 {\text{\ensuremath{R_{\textup{J}}}}}
\DeclareSIUnit[]\Re
 {\text{\ensuremath{R_{\oplus} } } } 
\DeclareSIUnit[]\Me
 {\text{\ensuremath{M_{\oplus} } } }
 
\begin{document} 


 \title{Investigating hot-Jupiter inflated radii with hierarchical Bayesian modelling}

 \author{
  Marko Sestovic\inst{1,2} 
  \and
  Brice-Olivier Demory\inst{1}
  \and
  Didier Queloz\inst{2,3}
  }
  
 \institute{
  Center for Space and Habitability, University of Bern, Gesellschaftsstrasse 6, Bern CH-3012, Switzerland\\
  \email{marko.sestovic@csh.unibe.ch}
  \and
  Astrophysics Group, Cavendish Laboratory, J.J. Thomson Avenue, Cambridge CB3 0HE, UK\\
  \and
  Observatoire de Geneve, Universite de Geneve, 51 chemin des Maillettes, CH-1290 Sauverny, Switzerland
  }

 \date{Received 28 June 2017 / 
 Accepted 27 March 2018}




\abstract
  {As of today, hundreds of hot Jupiters have been found, yet the inflated radii of a large fraction of them remain unexplained. A number of mechanisms have been proposed to explain these anomalous radii, however most can work only under certain conditions, and may not be enough to explain the most extreme cases. It is still unclear whether a single mechanism is enough to explain the entire distribution of radii, or whether a combination of them is needed.}
  {We seek to understand the relationship of radius with stellar irradiation and mass, and to find the range of masses over which hot Jupiters are inflated. We also aim to find the intrinsic physical scatter in their radii, caused by unobservable parameters, and to constrain the fraction of hot Jupiters that exhibit inflation.} 
  {By constructing a hierarchical Bayesian model, we infer the probabilistic relation between planet radius, mass and incident flux for a sample of 286 gas giants. We separately incorporate the observational uncertainties of the data and the intrinsic physical scatter in the population. This allows us to treat the intrinsic physical scatter in radii (due to latent parameters such as the heavy element fraction) as a parameter to be inferred.}%
  {We find that the planetary mass plays a key role in the inflation extent, with planets in the range $\sim 0.37-0.98\si{\Mj}$ showing the most inflated radii. At higher masses, the radius response to incident flux begins to decrease. Below a threshold of $0.37 \pm 0.03\si{\Mj}$ we find that giant exoplanets as a population are unable to maintain inflated radii $\gtrapprox 1.4\si{\Rj}$ but instead exhibit smaller sizes as the incident flux is increased beyond $10^6\si{Wm^{-2}}$. We also find that below $1\si{\Mj}$, there is a cutoff point at high incident flux beyond which we find no more inflated planets, and that this cutoff point decreases as the mass decreases. At incident fluxes higher than $\sim 1.6 \times 10^6 \si{Wm^{-2}}$ and in a mass range $0.37-0.98\si{\Mj}$, we find no evidence for a population of non-inflated hot Jupiters. Our study sheds a fresh light on one of the key questions in the field and demonstrates the importance of population-level analysis to grasp the underlying properties of exoplanets.}
  {}


 \maketitle




\section{Introduction}

Understanding the internal state and energy processes of giant planets is a key goal in exoplanet science. Gas giant interiors are generally modelled by assuming an adiabatic temperature profile and internal structure, and calculating the evolution of the planet radius as it contracts (for a review see, e.g. \cite{fortney_ssr2010, fortney_exobook2010, baraffe2014}).\par

 \cite{fortney2007,baraffe2008} modelled the evolution of planetary radii by including the effects of stellar irradiation and heavy element cores with state-of-the-art internal equations of state. Such model predictions can be tested against the currently observed population of exoplanets with known radii, masses and orbital distances. At low stellar irradiation, these models are thought to agree with observations, but the high-irradiation regime still presents a challenge. At fluxes greater than $\sim$\SI{2e5}{W.m^{-2}} \citep{miller2011,demory2011}, a sizeable fraction of gas giants are found to be anomalously larger than predicted, such as the  hot-Jupiters WASP-17 b, WASP-121 b and Kepler-435 b, which all have measured radii $R > 1.8\si{\Rj}$ \citep{anderson2011,almenara2015,delrez2016}.\par 
  
There are several candidate explanations for this inflation process, including tidal dissipation \citep{bodenheimer2001,bodenheimer2003,arras2010,jermyn2017}, kinetic heating \citep{guillot2002}, enhanced atmospheric opacities \citep{burrows2007}, double diffusive convection \citep{chabrier2007}, vertical advection of potential temperature \citep{youdin2010,tremblin2017}, and Ohmic heating through magnetohydrodynamic effects \citep{batygin2010,perna2010,wu2013,ginzburg2016}. \par 

 In most cases to date, theoretical predictions are tested against individual planets, or by producing radius predictions for a narrow set of parameters. However, the radii may depend on many parameters, including some that are non-observable or poorly constrained (latent parameters), such as the core mass, system age, internal composition and atmospheric opacity. The degeneracy between these parameters and the strength of the inflationary process limits the information we can obtain from a single planet. Thanks to the sheer count of exoplanets discovered so far, we are now in a position to study the issue of inflated radii using population-level analyses.
 
 
 In such an attempt, the choice of forward model is central. The candidate model should not only reproduce a range of individual planet observations, but also predict the full distribution of the observable planet radii and their dependence on planet mass and stellar incident flux, within reasonable assumptions on the latent parameters. \cite{laughlin2011} already used the observed radii of 90 well-characterised transiting exoplanets to assess the incidence of Ohmic heating on the inflated radius anomaly (first introduced by \cite{guillot2006}). Using a population-level approach, one can also extract model-independent conclusions from the data. For example \cite{enoch2012} use multivariate regression models to investigate the factors that affect the radii of 119 extrasolar gas planets, such as planetary equilibrium temperature, planet mass, stellar metallicity and tidal heating rate.\par
 
 To that end, it is important to separate the real physical scatter caused by the latent parameters, and the scatter caused by measurement uncertainties (see \cite{wolfgang2016} where this was addressed for super-Earths and sub-Neptunes). For example, hot Jupiter radii span from $\sim1.99\si{\Rj}$ (Kepler-435 b, \cite{delrez2016}) to $\sim0.84\si{\Rj}$ (Kepler-41 b, \cite{santerne2011}). We seek to understand how much of this spread can be accounted for by the measurement uncertainty in the radius, and thus constrain the ``true'' range of radii that is driven by the currently unidentified physical process.\par  
 
 In this work, we infer the flux-mass-radius (FMR) relation using a population of 286 transiting gas giants with measured masses and radii. 
 We use a very similar approach to e.g. \cite{wolfgang2016,shabram2015} by employing a hierarchical Bayesian model (HBM) in which we separate the effects of intrinsic scatter and measurement uncertainty \citep[see also][]{kelly2007,hogg2010a,demory2014,rogers2015}. This also allows us to include the uncertainty in both the dependent parameter (planet radius) and the independent parameters (planet mass and incident flux), which cannot be done with basic least-squares regression. We thus constrain the FMR relation as a distribution rather than a deterministic function, and characterise the physical scatter of the gas giant population. \par

 It is as of yet unknown whether all hot Jupiters are inflated, or if some may have radii that agree with non-inflationary models \citep[i.e][]{fortney2007}. Many of the candidate inflation mechanisms would not apply for all planets, e.g. tidal heating for circularised orbits \citep{gu2003}. Similarly, \cite{heng2012,perna2012} showed that the depth and degree of Ohmic heating is sensitive to numerous atmospheric parameters such as opacity or thermal inversions and may be inhibited in some cases. Measuring the FMR distribution and its intrinsic scatter thus provides a way to determine whether such non-anomalous planets exist.\par
 
 We present the planet data used in this study in Sect. \ref{data} and present our parameterisation of the FMR distribution in Sect. \ref{model}. The methods for fitting the parameters and selecting our model are explained in Sect. \ref{hbm}, and the results of our fit for the FMR distribution are shown in Sect. \ref{results}. We test the agreement of our model with the data in Sect. \ref{check}, and we finally discuss the implications of our results in Sect. \ref{discussion}. \par

\begin{figure}
 \centering
 \includegraphics[scale=0.5]{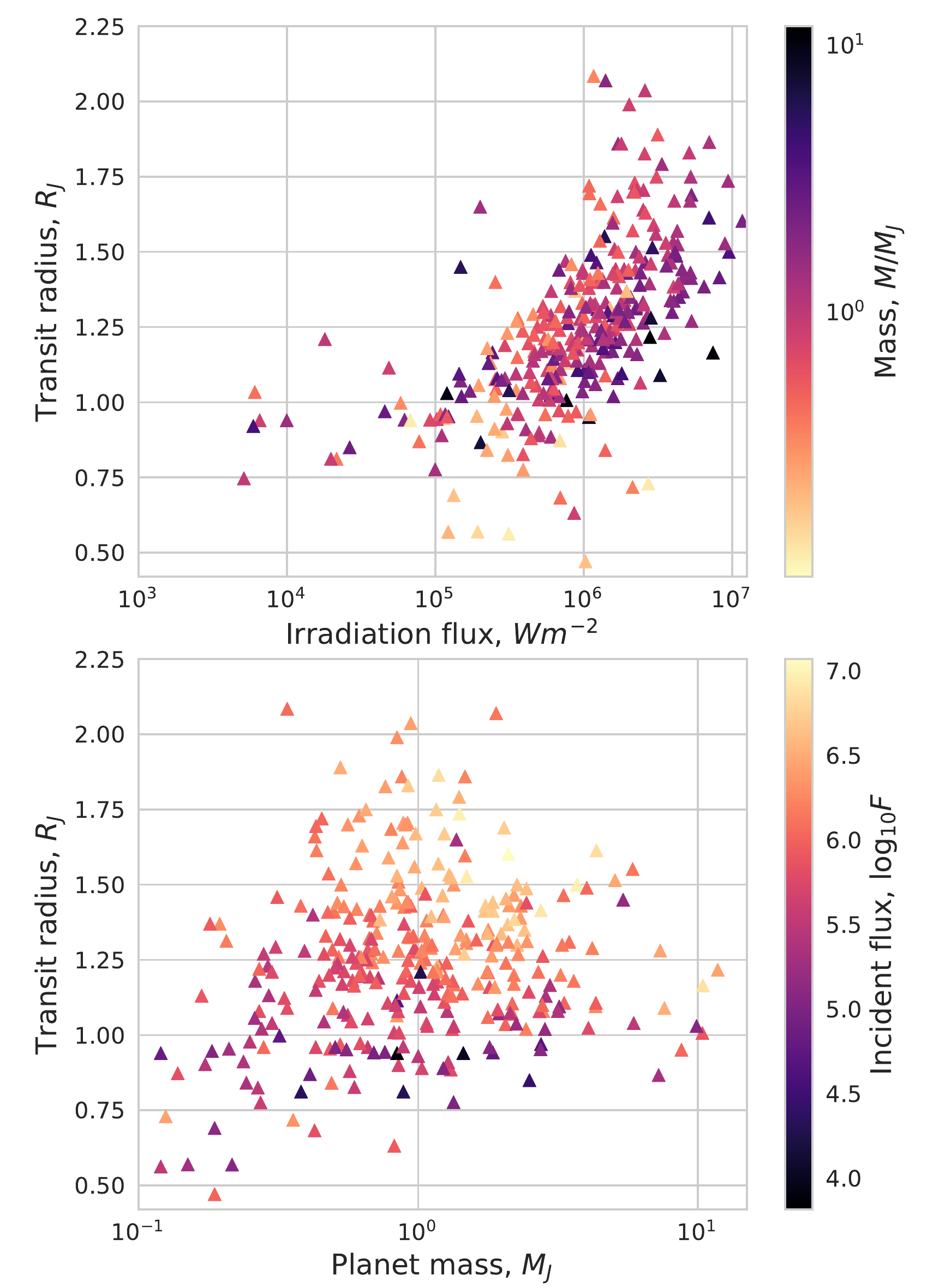}
 \caption{Top: the distribution of the transit radii in our sample vs the incident flux, with the mass colour-coded. The radii begin to show a correlation with flux at $\sim$\SI{2e5}{W.m^{-2}}, corresponding to where inflated hot Jupiters begin to appear \citep{demory2011}. Bottom: the transit radii plotted against mass, colour coded by incident flux.}
 \label{preplot}
\end{figure}


\section{Data}
\label{data}

As we require both the mass and the radius of the planets, we include only transiting giants that have radial velocity (RV) or transit timing variation (TTV) mass measurements. Of  those, 9 planets have masses derived from TTVs, which are biased towards lower fluxes and masses, with uncertainties that are larger on average than for the whole sample\footnote{The average fractional uncertainty is $8.8\%$ for the RV masses, and $15.0\%$ for the TTV masses.}. We include them because they populate a relatively sparse region of the parameter space, however we note that they may have a different selection bias than the rest of the sample (see e.g. \cite{wolfgang2016}, where such a bias was found for sub-Neptune planets). We also use the stellar temperature, stellar radius, and the semi-major axis of the planetary orbit to calculate a distribution for the incident flux on the planet (see Sect. \ref{hbm}). 

The data is taken from the Exoplanet Orbit Database \citep{eod} and supplemented by the NASA Exoplanet Archive \citep{akeson2013}, both last accessed on 10/10/2016. We check the Archive and the source paper in cases where the EOD data is incomplete. Planets that are missing required values in both databases are discarded. We also take information about the planet ages and metallicities from the exoplanet.eu database \citep{schneider2011}. Where the raw data has asymmetric error bars, we average them and take the uncertainty distribution to be Gaussian. This follows the treatment of \cite{weiss2013,wolfgang2016}, since we do not have access to the full likelihood distributions for all our parameters. While this may introduce a bias into our results, a short test (see Sect. \ref{trend}) suggests the effect on our hyperparameters is well within our posterior distribution uncertainties.

Our main criterion for selecting our sample is the planet mass.  We set $13\si{\Mj}$ as the upper cutoff to exclude planets where deuterium burning is thought to happen \citep{chabrier2005,spiegel2011}. Choosing the lower bound is related to the wider problem of whether there are any strong boundaries in gas-giant characteristics, which is one of the questions we seek to answer. \cite{enoch2012} chose $0.1-0.5\si{\Mj}$ to define Saturn-mass planets and $0.5-2.0\si{\Mj}$ for Jupiter-like planets, finding that Saturn-mass planets do not exhibit any inflation with temperature. \cite{laughlin2011} also use $0.1\si{\Mj}$ as a lower bound for their sample.

We choose $0.1\si{\Mj}$ as an absolute lower bound, with the understanding that this will introduce planets into our model that do not have the same properties as the rest of our sample, i.e. Saturn and Neptune-like planets which may not exhibit inflation, as shown by \cite{enoch2012}. Finding mass boundaries between the hot-Jupiter planets that inflate and those that do not is part of our HBM implementation (see Sects. \ref{model} and \ref{hbm}).


\section{Forward Model}
\label{model}

In this section, we review the known trends and present our main parametrisation for the flux-mass-radius distribution. We will refer to the following as the \emph{baseline} model.

Our current understanding of giant-planet structure and evolution is that at low incident flux, thermal evolution models can reproduce the range of planet radii present in the exoplanet population \citep{miller2011,thorngren2016}. Although previous models predict that radii will depend on stellar irradiation, particularly at fluxes of $\sim$\SI{1e5}{W.m^{-2}} and greater ($\sim$0.1 AU), it is also clear that a large fraction of gas giants in short-period orbits are much more inflated than predicted \citep[see][]{fortney2007,baraffe2008,fortney_ssr2010, fortney_exobook2010, baraffe2014}. \cite{demory2011,miller2011} showed that these inflated hot Jupiters start to occur at incident fluxes of greater than $\sim$\SI{2e5}{W.m^{-2}}. \cite{enoch2012} and \cite{laughlin2011} also find that the radii of these hot gas giants can depend strongly on irradiation temperature. These trends can be seen in the top of Fig. \ref{preplot}.\par

Taking this into account when choosing the parametrisation of our empirical flux-mass-radius relation (represented as the forward model for radius, $R$, given mass and flux, $M$ and $F$), we model the effect of stellar irradiation by considering two regimes. For planets below a threshold $F_s$ in incident flux, the radii are taken to be constant at size $C$, and independent of flux. Above $F_s$ the radii increase proportionally with $\log F$. This relation (Equation \ref{deterministic}) qualitatively agrees with the distribution of radii and fluxes in Fig. \ref{preplot}:

\begin{equation}
 \frac{R}{\si{\Rj}}=\begin{cases}
 C, & F<F_{s}\\
 C+A\cdot (\log F-\log F_{s}), & F\geq F_{s}
 \end{cases}
 \label{deterministic}
\end{equation}

While planetary radii may depend on incident flux even at $F < F_s$, this dependence is very weak, and variations in age, core mass and envelope composition are expected to play a significantly greater role, see e.g. \cite{fortney2007,baraffe2008,miller2011}. Additionally, due to a lack of transiting planets with known masses found at large orbital distances (there are only $22$ planets further out than $0.1\si{AU}$ in our sample), combined with the observational uncertainty of the measurements, it is unlikely that we could extract meaningful constraints about the dependence of planetary radius with incident flux for gas giants at low irradiation. \par

Equation \ref{deterministic} provides a deterministic way of modelling the planetary radius based on the incident flux and the model parameters which we must infer ($A$, $C$ and $F_s$). However due to non-observable parameters, such as core mass, composition and age, we expect to see a physical scatter in the radii at a given flux level. We include this uncertainty by attaching to the planet radius a Gaussian distribution of standard deviation $\sigma_R$, around a mean value $\mu_{R}$ which is given the same form as Equation \ref{deterministic}:

\begin{equation} 
 R \sim N(\mu = \mu_R \, ,\, \sigma = \sigma_R)
 \label{R}
\end{equation}

\begin{equation}
\mu _{R}(F,M)=\begin{cases}
C(M), & F<F_{s}(M)\\
C(M)+A(M)\cdot\log_{10} \frac{F}{F_{s}(M)}, & F\geq F_{s}(M)
\end{cases}
\label{mu}
\end{equation}


where $\sim N(\mu, \sigma)$ means `drawn from a normal distribution with mean $\mu$ and variance $\sigma^2$.' \cite{enoch2012} found that there were large differences in the radius-temperature relations between planets of different masses. We thus include the effect of planet mass in our model by splitting our total mass range into $J$ separate bins, each with a different radius-flux relation of the same form as Equation \ref{R}, but with independent model parameters ($C$, $F_s$ and $A$):

\begin{equation}
A(M)=\begin{cases}
A_{1}, & \frac{M}{\si{\Mj}}<m_1\\
... \\
A_{j}, & m_{j-1} \leq \frac{M}{\si{\Mj}} < m_j\\
... \\
A_{J}, & m_{J-1} \leq \frac{M}{\si{\Mj}}
\end{cases}
\label{A}
\end{equation}

with $C(M)$ and $F_s(M)$ following the same form. $\left\{m_j\right\}^{J-1} _{j=1}$ represent the inner boundaries separating the mass bins. Unlike previous studies, we do not fix the mass boundaries to be constant, as their placement would be arbitrary and could bias the FMR relation. Instead, the mass bin boundaries are also model parameters that we infer. \par

Our final baseline model uses a mass- and flux-dependent scatter, $\sigma_{R}(M,F)$, with three possible values for three classes of planets: planets in the lowest mass bin ($M < m_1$), and inflated and weakly-irradiated planets ($F \geqslant F_s$ and $F<F_s$ respectively) shared for all higher mass bins ($M \geqslant m_1$):

\begin{equation}
\sigma_{R}(F,M)=\begin{cases}
\sigma_{R,1}, & \frac{M}{\si{\Mj}}<m_1\\
\sigma_{R,2}, & \frac{M}{\si{\Mj}}\geq m_1 \text{ and } F<F_{s}\\
\sigma_{R,3}, & \frac{M}{\si{\Mj}}\geq m_1 \text{ and } F\geq F_{s}
\end{cases}
\label{scatter_halfsplit}
\end{equation}

The baseline model, with 4 mass bins and 17 parameters, was chosen after a process of model selection, where we considered several alternative parametrisations and compared their posterior results. We varied the number of mass bins, choosing between 3, 4 and 5 bins, as well as the form of $\sigma_R$ and $F_s$. For the flux threshold, we compared with the case where the threshold was the same for all mass bins. For the intrinsic scatter, we also set it to depend on just the mass instead, or on both mass and incident flux, and we tried models where the intrinsic scatter was the same for all fluxes and masses (as a constant single parameter). We did this by considering which models converged to unimodal posteriors, and with tests on our posterior predictive fit, described in Sect. \ref{select_tools} and Sect. \ref{check}.


\section{Hierarchical Bayesian model}
\label{hbm}

\begin{figure}
 \includegraphics[scale=0.6]{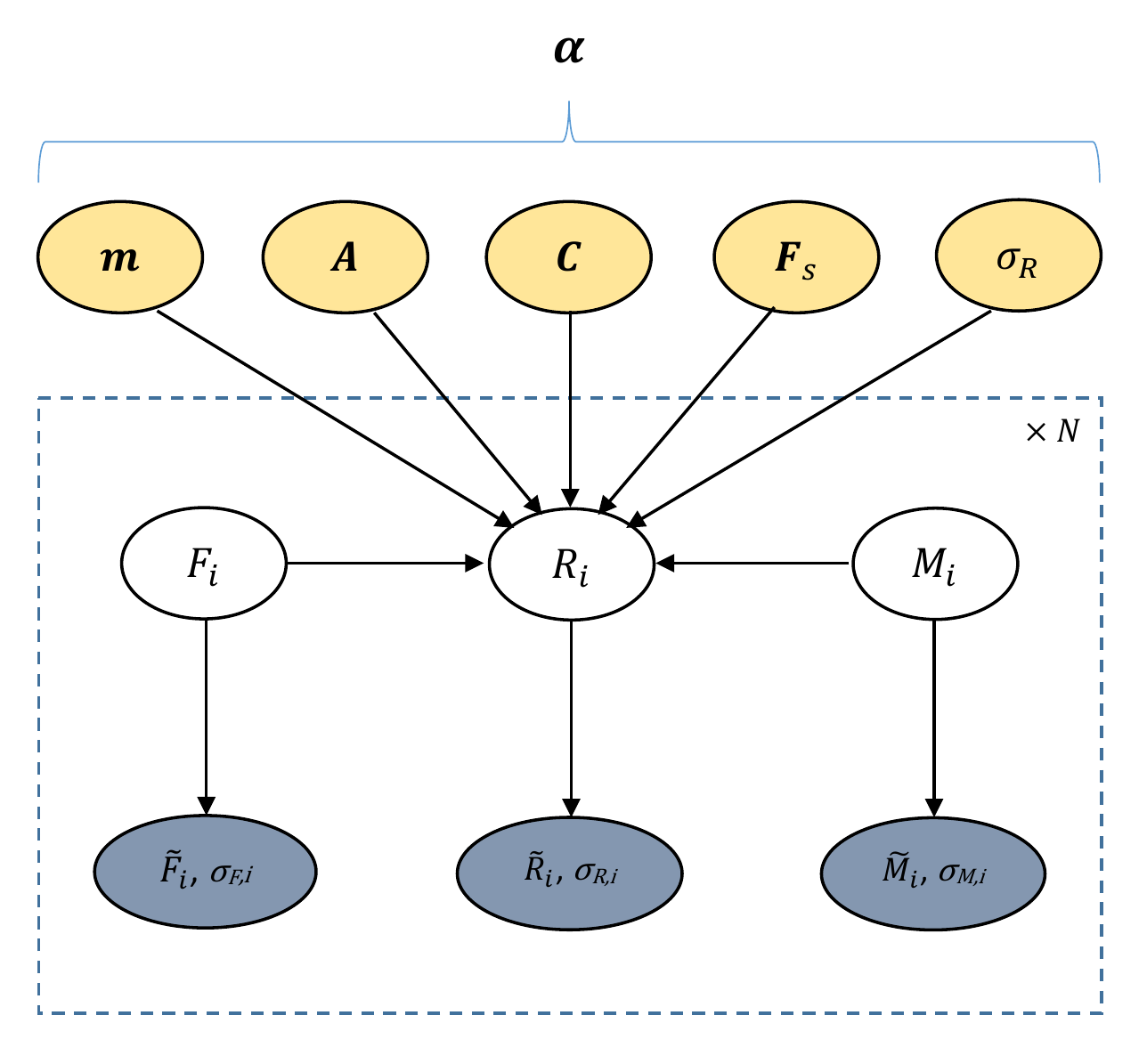}
 \caption{Our model shown as a Bayesian network, with the forward model hyperparameters in light orange (the third level of the HBM), the "true" planet parameters in white (the second level of the HBM), and the fixed observed values in dark blue. The arrows denote conditional dependencies, with the probability distribution of a "child" parameter dependent on the values of its "parent" parameters (arrows go from parent to child).}
 \label{bayesnetwork}
\end{figure}

Our aim is to infer the values of the model parameters\footnote{In this context, known as the hyperparameters.} of the distribution for R given F and M, or $R|F,M$. We will also refer to the parameters by shorthand such as $\mathbf{A}=\left\{A_j\right\}^J _{j=0}$, or all of them together as $\boldsymbol{\alpha} = (\mathbf{A},\mathbf{C},\mathbf{F}_s,\mathbf{m},\boldsymbol{\sigma}_R)$ . These are not necessarily vectors however. \par


The probabilistic model described in Sect. \ref{model} defines the relationship between the true parameters $R$, $M$ and $F$ of a planet. However when an observation is made, e.g. measuring the transit radius $\widetilde{R}$, the result is subject to noise. It can be taken as being drawn from a probability distribution which depends on the ``true'' radius $R$ and the noise characteristics of the observation. This is represented as $\widetilde{R}\sim p(\widetilde{R}|R,\boldsymbol{\alpha}_{obs})$, where $\boldsymbol{\alpha}_{obs}$ parametrize the noise of a particular observation. We take care to differentiate between the observed value $\widetilde{R}$ and the true value $R$. In the most simple case, $\widetilde{R}$ is drawn from a Gaussian distribution with mean $R$ and a standard deviation which is the observational uncertainty, $\sigma_{obs}$. As shorthand, we will refer to the combined observational data as $\tilde{\boldsymbol{x}} = \left\{\widetilde{R}_i,\widetilde{M}_i,\widetilde{F}_i\right\}_{i=1} ^N$.\par

The distribution of our observables $(\widetilde{R},\widetilde{M},\widetilde{F})$ for a given planet is the distribution of the true parameters $(R,M,F)$, convolved with the observational uncertainty distribution. Thus the true scatter of the radii, $\sigma_{R}$, is lower than the scatter of our observed data, and the posterior values for our parameters $(R,M,F)$ will be less spread out than the data (see Figs. \ref{multirelation_data} and \ref{multirelation_post}). This is a general consequence in cases where quantities drawn from the same distribution (the true values) are subject to an additional scatter, see \cite{stein1956} and \cite{good1965}.\par


Treating this problem in a hierarchical Bayesian model (HBM) framework allows us to separately incorporate both the measurement errors in all the parameters, and the intrinsic scatter in the radius (which is a parameter to be inferred). We refer the reader to e.g. \cite{wolfgang2016} for a brief overview of the advantages of HBMs for a very similar problem. By defining the probabilistic relations and placing priors on all the parameters, we can calculate the joint posterior distribution for $\mathbf{R}$, $\mathbf{F}$, $\mathbf{M}$, and $\boldsymbol{\alpha}$, given the data. The forward model relation for $R_i | M_i,F_i,\boldsymbol{\alpha}$ has already been defined in Sect. \ref{model}, here we define the other probabilistic relations and priors (with their dependence relationships shown in Fig. \ref{bayesnetwork}):

\begin{eqnarray*}
2\log _e\sigma_{R,j} & \sim & U(-10,4)\\
m_j & \sim & U(m_{j-i},m_{j+1})\\
\log_{10} F_{s,j} & \sim & N(5.3,1)\\
C_{j} & \sim & U(0,2)\\
\arctan A_{j} & \sim & U(-\frac{\pi}{2} ,\frac{\pi}{2})\\
M_{i} & \sim & U(0,\infty)\\
\widetilde{R}_{i} & \sim & N(R_{i},\sigma_{R,obs,i})\\
\widetilde{M}_{i} & \sim & N(M_{i},\sigma_{M,obs,i})\\
F_{i} & \sim & SN(\mu_{F,i},\sigma_{F,obs,i},\alpha_{F,i}) \\
R_i | M_i,F_i,\boldsymbol{\alpha} & \sim & N(\mu_R(M_i, F_i, \boldsymbol{\alpha}),\sigma_R)
\end{eqnarray*}

where $U(a,b)$ represents a uniform distribution with upper and lower bounds $a$ and $b$ respectively, and $SN(\mu,\sigma,\alpha)$ represents a skew-normal with the skewness parameter $\alpha$ \citep{azzalini}.\par

We use an uninformative prior for $A_j$ (a Jeffreys prior for the slope of a line, see \cite{vanderplas2014}) and essentially uninformative priors for $\sigma_{R,j}$ and $C_j$. The cutoffs on $\sigma_{R,j}$ and $C_j$ are placed at values beyond which we can assume to have negligible likelihood, though this is arbitrary. An average cold radius of greater than $2\si{\Rj}$ or less than zero is clearly not supported by the data, and a scatter of less than $0.007\si{\Rj}$ is much less than expected based on possible variations in core mass and planet age. The posteriors are not found to have any significant probability density near these limits, so our choice is justified.\par


The mass bin boundaries are given a uniform prior but constrained to prevent them from crossing each other. There are no planets with mass less than $0.1\si{\Mj}$ in our sample (see Sect. \ref{data}), thus we place a minimum cutoff for the lowest bound at $0.1\si{\Mj}$. Among the model parametrisations that are tried, we find that in models with 4 bins (i.e. 3 boundaries), the highest boundary generally has a very wide distribution and would occasionally get stuck outside our mass range. However, we also find that when we fix the highest mass bin boundary to an arbitrary value larger than $2.0\si{\Mj}$, we still obtain clearly distinct behaviour in each of the four mass bins (in terms of the radius response to inflation). In the interest of extracting the maximum amount of information from the data, we thus use 4 mass bins, but with the highest boundary ($m_3$) fixed at $2.5\si{\Mj}$ (see Section \ref{select_tools}).


The prior for $F_s$ is based on the results of \cite{demory2011,miller2011}, which found that hot-Jupiter inflation is no longer obvious below an incident flux of $\sim$\SI{2e5}{W.m^{-2}}. This constraint is only weakly informative, with a conservative $1\sigma$ interval between \SI{2e4}{} and \SI{2e6}{W.m^{-2}}.\par

One of the main benefits of Bayesian techniques is the ability to set priors based on known physical constraints (for example a maximum density based on a fully iron giant planet). However in this case, we do not insert any physical priors. According to \cite{fortney2007} radii of $0.5-0.6\si{\Rj}$ are possible for low mass giants with heavy cores in a 50-50 rock/ice mixture, and even smaller radii would be possible for pure rock and iron planets (down to less than $0.3\si{\Rj}$), which is significantly lower than any observed radius in our sample. Thus, it is unlikely that adding a maximum density constraint would have affected our results. We also prefer not to insert constraints from interior models into our HBM to keep it purely observation driven, so we simply constrain the true mass, radii and fluxes to be positive. Apart from the constraints on the mass bin boundaries, we do not find that our choice of priors significantly affects the posterior distributions for any of the other parameters.\par

The incident flux depends on the stellar effective temperature, orbital distance, and stellar radius as:

\begin{equation}
    F =  \frac{R_{\star}^2}{a^2} \sigma T_{\star}^4
\end{equation}

where $\sigma$ is the Stefan-Boltzmann constant, $a$ is the orbital distance (taken as the semi-major axis), and $R_\star$ and $T_\star$ are the stellar radius and temperature respectively. For each planet, we calculate the distribution of flux based on those parameters, before solving the HBM. We find that the skew-normal distribution, described in \cite{azzalini}, is a very good fit to the resulting asymmetric distributions of flux. We therefore fit a skew-normal distribution for each planet's incident flux, and extract the best-fit parameters $\mu$, $\sigma$ and the skewness $\alpha$. This reduces our number of free parameters by nearly a factor of two, aiding in convergence time. \par

Our data come from a large range of programs and sources, and we do not include the statistical biases introduced during observation. Observations currently favour planets with large radii and masses, however the nature of this bias depends on the instrument, and how close a parameter is to the detection threshold. The statistical biases in the flux parameter are also complicated; higher fluxes are favoured due to closer orbital distances, but disfavoured due to a smaller transit depth for larger stars. We must take this into consideration when forming our conclusions. There are also selection effects in the ground-based RV follow-up that we are unable to model.\par

We find the posterior distributions of our parameters by drawing samples from the joint posterior distribution $p(\boldsymbol{R},\boldsymbol{F},\boldsymbol{M},\boldsymbol{\alpha}|\tilde{\boldsymbol{x}})$, using a Metropolis-Hastings sampler \citep{hastings1970,chib_greenberg1995} implemented in the PyMC2 package in python \citep{pymc}. Hamiltonian Monte Carlo samplers such the No-U-Turn Sampler \citep{hoffman2011}, implemented in PyMC3, cannot be implemented due to the discontinuous nature of our forward model. Other alternatives, such as the affine invariant ensemble sampler, implemented in emcee, are unsuitable due to the nearly 900 dimensions in our model \citep{goodman2010,emcee}.


\subsection{Model selection tools}
\label{select_tools}


While increasing the number of hyperparameters can lead to a better fit to the data, it can also lead to overfitting. Increasing the model complexity gives our model more freedom, which can lead to multi-modal posterior distributions which our M-H sampler can't accurately sample, leading to un-converged posteriors. We fit multiple parametrisations, and base our final choice firstly on whether it could converge and whether it has well defined uni-modal posteriors. To check for convergence, we run multiple MCMC chains, and calculate the Gelman-Rubin convergence metric for our posterior samples, where values close to 1 indicate convergence and good mixing of the chains \citep{gelman1992}.

To choose between converged models, we use two tools. Firstly, we calculate the Bayesian Information Criterion \citep{bic}, calculated from

\begin{equation}
BIC = k\ln N - 2\ln \hat{L}
\end{equation} 

where $k$ is the number of hyperparameters, $N$ is the number of samples (the number of planets), and the highest likelihood $\hat{L}$ is defined as:

\begin{eqnarray}
\hat{L} & = & p(\tilde{\boldsymbol{x}}|\hat{\boldsymbol{M}},\hat{\boldsymbol{F}},\hat{\boldsymbol{\alpha}},\mathcal{M})\\
 & = & \int p(\tilde{\boldsymbol{x}}|\boldsymbol{R},\hat{\boldsymbol{M}},\hat{\boldsymbol{F}})p(\boldsymbol{R}|\hat{\boldsymbol{M}},\hat{\boldsymbol{F}},\hat{\boldsymbol{\alpha}},\mathcal{M})d\boldsymbol{R}\\
 & = & \prod _{i=1} ^{N} \int p(\tilde{\boldsymbol{x}}_i|R_i,\hat{M}_i,\hat{F}_i)p(R_i|\hat{M}_i,\hat{F}_i,\hat{\boldsymbol{\alpha}},\mathcal{M})dR_i
 \label{bic}
\end{eqnarray}
 
where by marginalising out the second level of our model, we obtain a two-level model, and equation \ref{bic} relies on the independence between observations. This is an assumption that we have to make as we cannot model the correlation between observations and between parameters with the publicly available information. $\mathcal{M}$ refers to the model in question, and $\hat{M}_i$, $\hat{F}_i$, $\hat{\boldsymbol{\alpha}}$ represent the model parameters set to their maximum likelihood values. A lower BIC indicates a better model, with differences of $\sim 10$ required for strong evidence in favour of a model \citep[see][]{bic}. Due to the Gaussian approximation required for the BIC, which is not satisfied for all our posterior parameter distributions, the BIC only plays a minor role in our model selection. We only consider very large differences (more than $~20$ in the BIC) to be significant.

A better tool for model testing is to perform detailed statistical tests on our posterior predictive fit. We simulate the data that would be produced by our model, and compare it to the real observed data. This essentially tests the question: ``if our fully marginalised model was correct, how likely would we be to produce the observed data?'' This model checking follows the procedure in \cite{wolfgang2016}, and is described in Sect. \ref{check}.

For the final baseline model, we decided to fix the highest mass boundary at $2.5\si{\Mj}$. In models where the highest mass boundary was treated as an unknown parameter, its posterior distribution was very wide and poorly constrained, generally varying between 2 and 10 Jupiter masses. This is likely due in part to the sparsity of such high mass planets in the database. Fixing it in place greatly sped up convergence times, and allowed us to focus on the lower masses where there is greater variation.


\begin{figure*}[t]
  \centering
  \begin{subfigure}{.49\textwidth}
    \centering
    \includegraphics[scale=0.48]{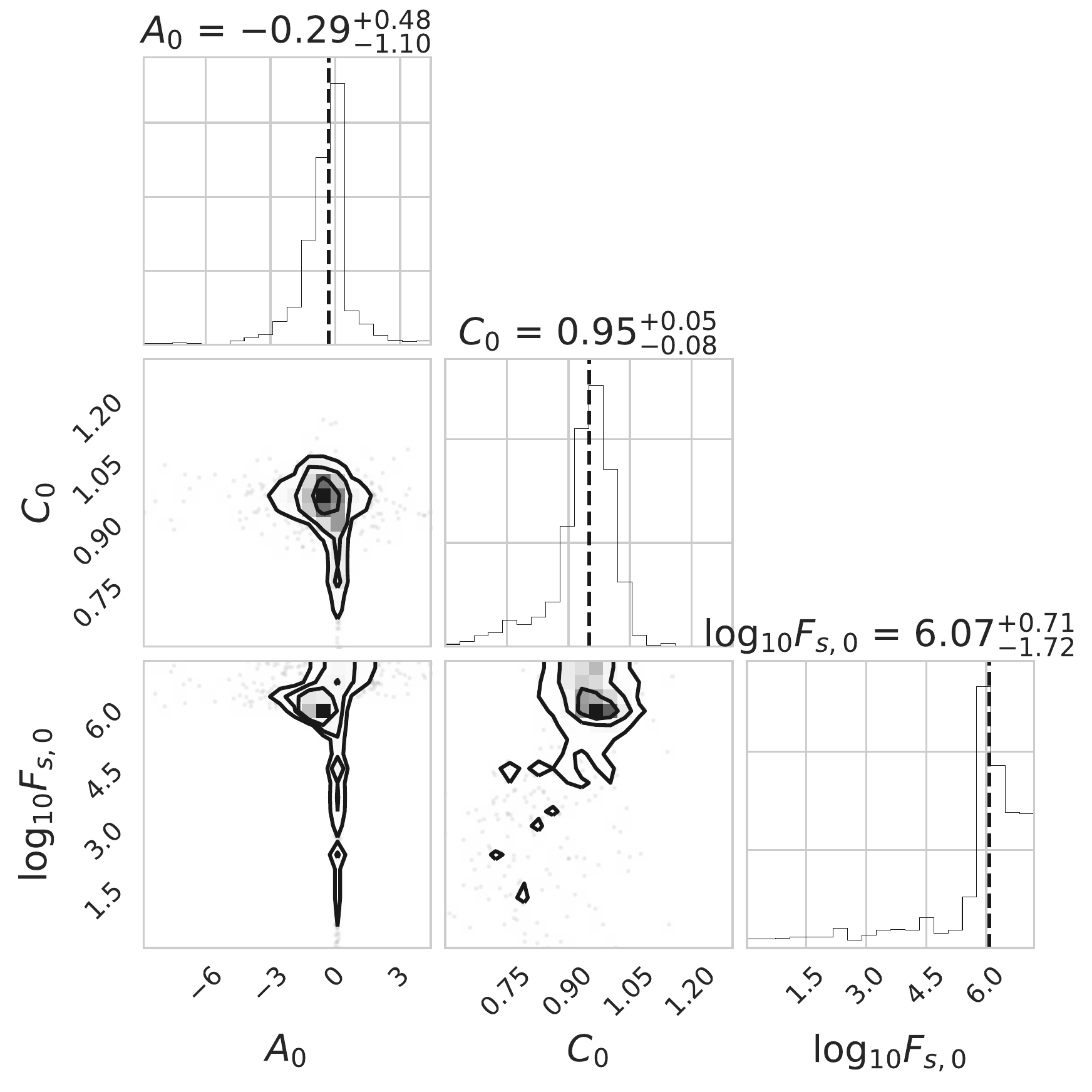}
    \caption{\ref{fig:HP}a: $0.1-0.37\si{\Mj}$}
    \label{fig:cp0}
  \end{subfigure}
  \hfill
  \begin{subfigure}{.49\textwidth}
    \centering
    \includegraphics[scale=0.48]{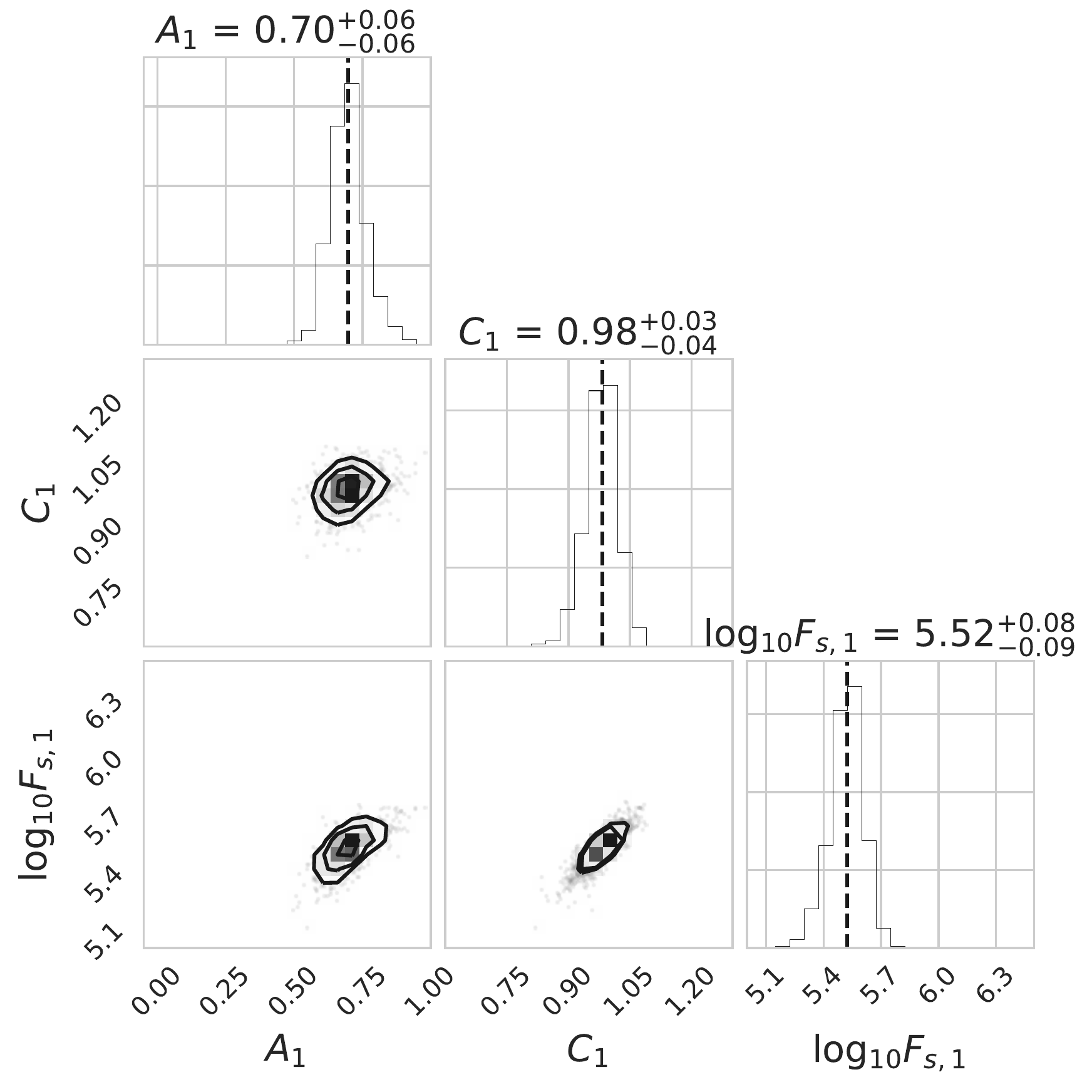}
    \caption{\ref{fig:HP}b: $0.37-0.98\si{\Mj}$}
    \label{fig:cp1}
  \end{subfigure}
  
  \begin{subfigure}{.49\textwidth}
    \centering
    \includegraphics[scale=0.48]{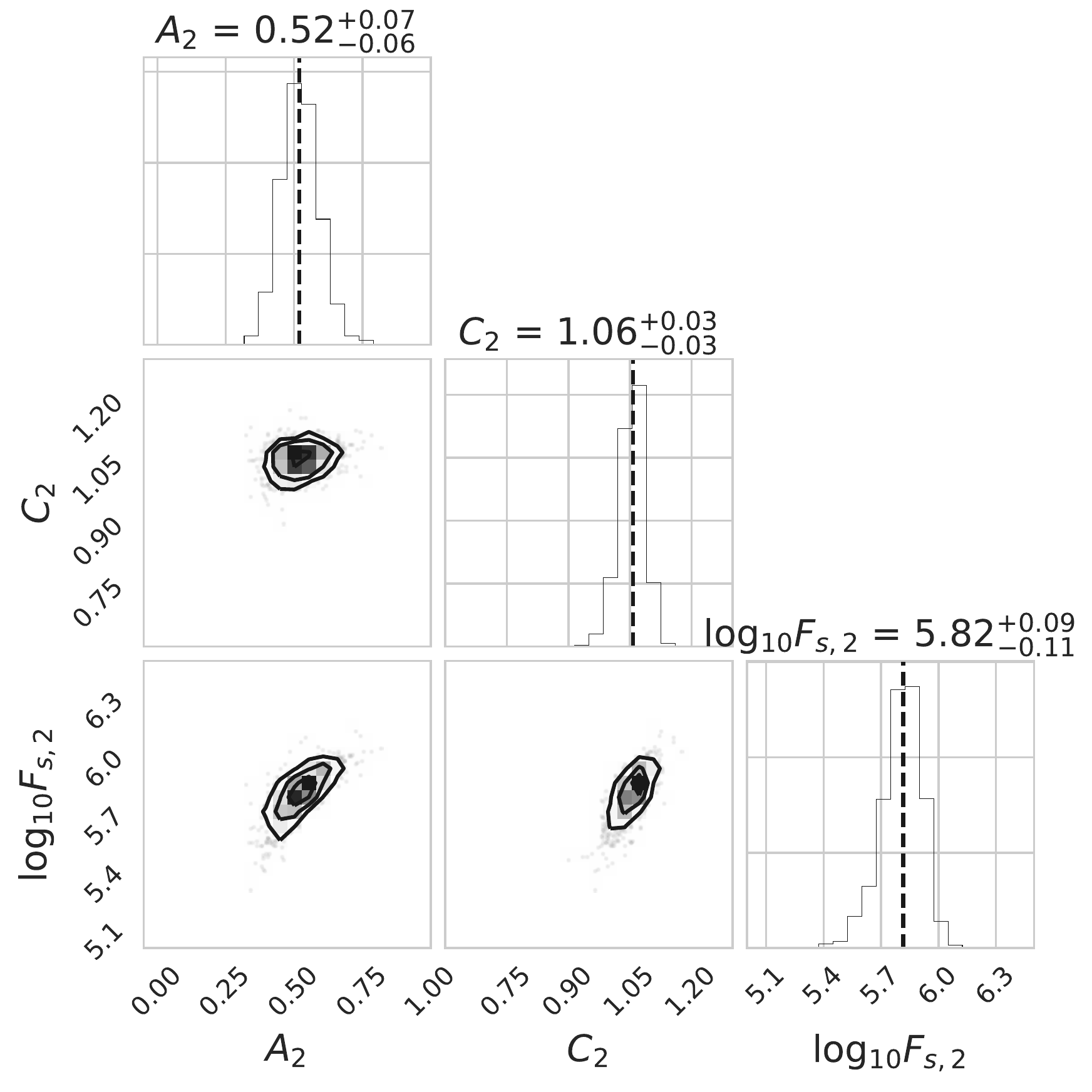}
    \caption{\ref{fig:HP}c: $0.98-2.50\si{\Mj}$}
    \label{fig:cp2}
  \end{subfigure}
  \hfill
  \begin{subfigure}{.49\textwidth}
    \centering
    \includegraphics[scale=0.48]{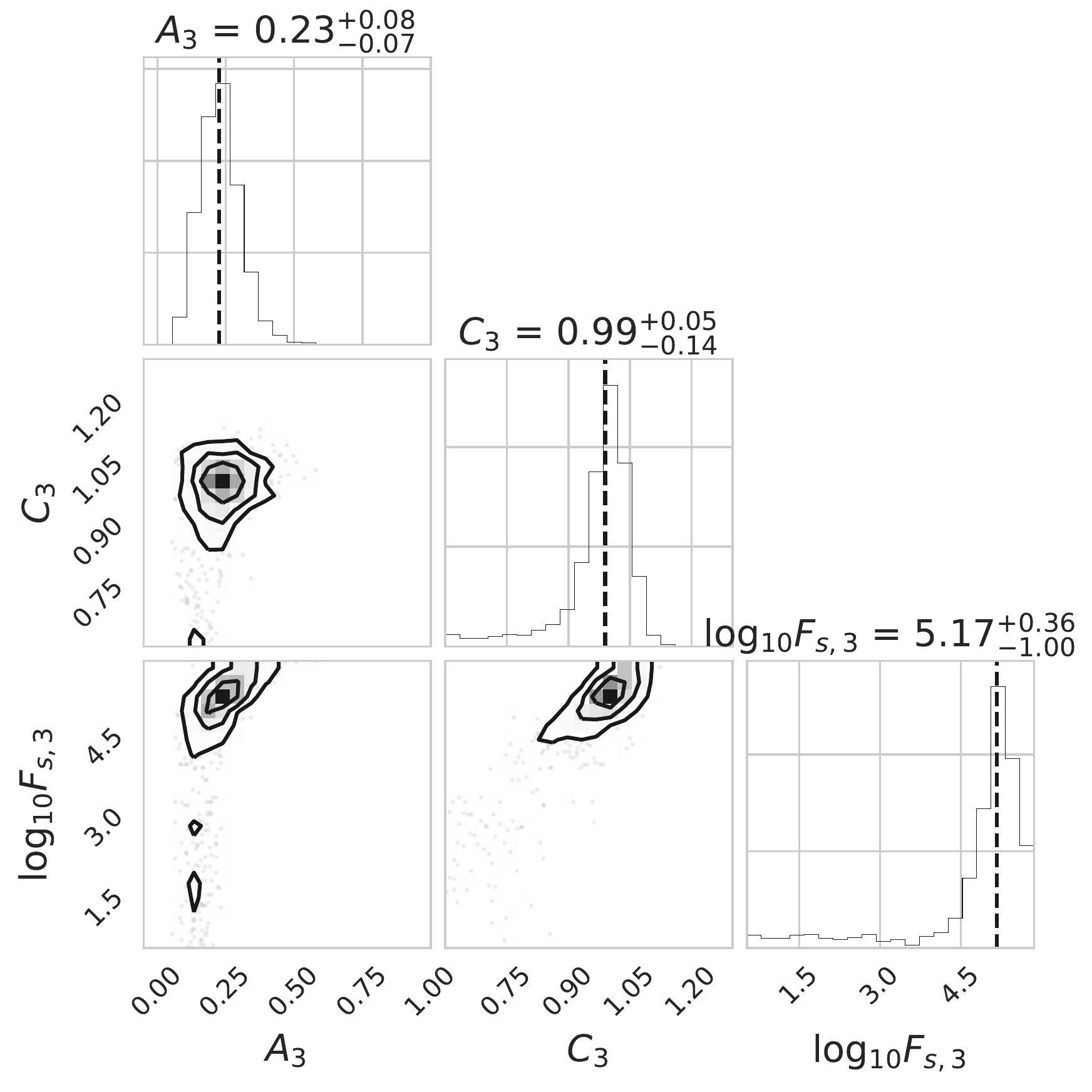}
    \caption{\ref{fig:HP}d: $2.50+\si{\Mj}$}
    \label{fig:cp3}
  \end{subfigure}
  
  \caption{The posterior distributions of the $A$, $C$ and $\log F_s$ parameters, for each mass regime.
  }
  \label{fig:HP}
\end{figure*}

\section{Results}
\label{results}

\subsection{Model posteriors}


We ran 10 chains for 2,000,000 iterations each, with a burn-in of 1,000,000 iterations, and a thin factor of 200. The Rubin-Gelman convergence metric was evaluated for the posterior distributions, and we achieved $\hat{R}<1.003$ for all the model parameters, indicating a high level of convergence \citep{gelman1992}.

\begin{figure}
 \centering
 \includegraphics[scale=0.5]{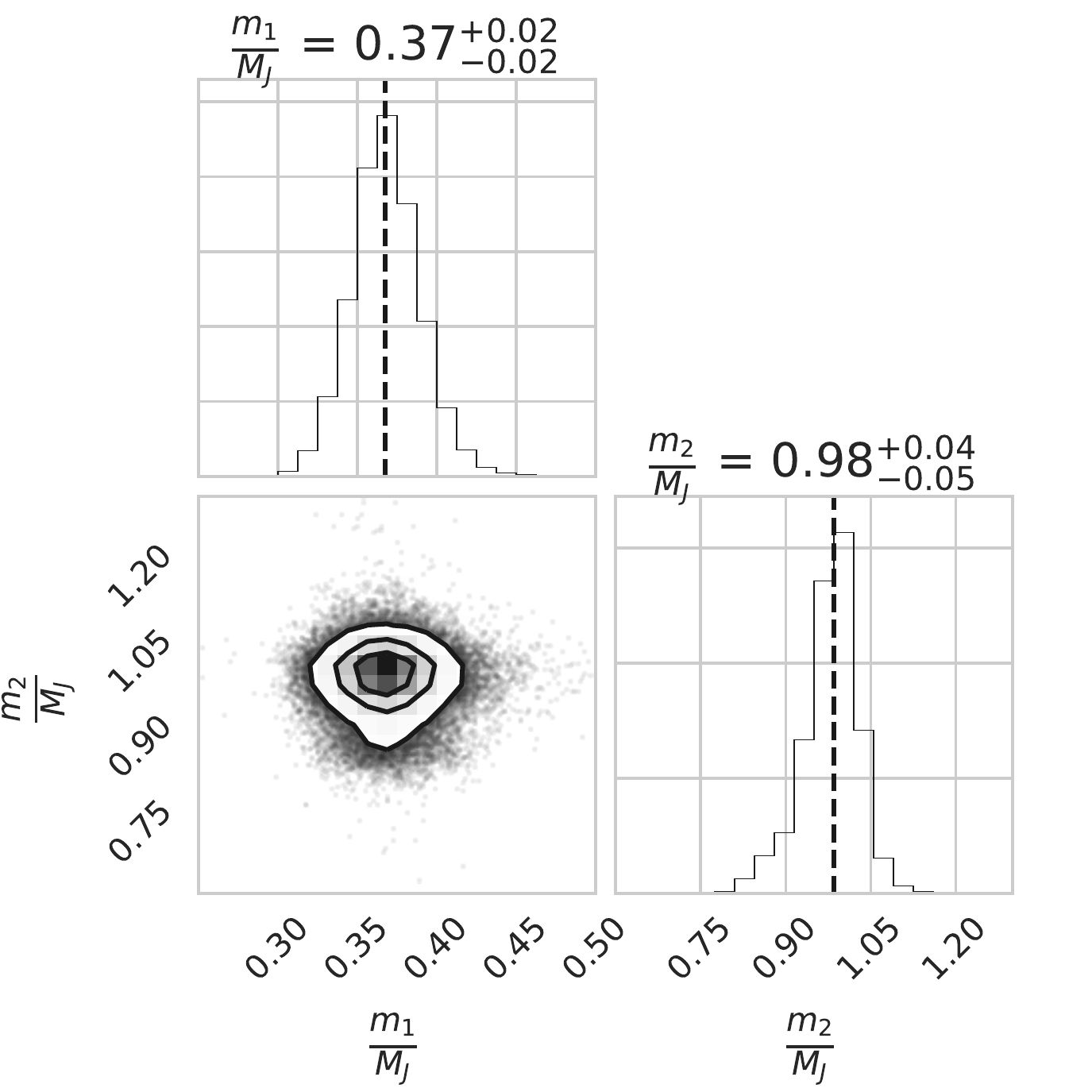}
 \caption{The posterior distribution of the two variable mass bin boundaries, $m_1$ and $m_2$. We kept $m_3$ fixed at $2.5\si{\Mj}$.}
 \label{MB}
\end{figure}

The posterior distributions for our mass bin boundaries, $m_1$ and $m_2$ (see Fig. \ref{MB}), and the fixed boundary at $m_3=2.5$, split our sample into four mass regimes: planets with sub-Saturn masses ($<0.37\si{\Mj}$), planets with sub-Jupiter masses ($0.37-0.98\si{\Mj}$), and planets heavier than Jupiter ($0.98-2.5\si{\Mj}$ and $2.5+\si{\Mj}$). Above the inflation threshold, the radii have a clear dependence on incident flux for all but the sub-Saturn mass planets, and the flux-radius relation for each mass bin is distinct (in the sense that the joint posterior distributions for the hyperparameters that govern it, $A$, $C$ and $F_s$, have little overlap, see Fig. \ref{fig:HP}).

We plot the posteriors for $C$, $A$ and $F_s$ in Fig. \ref{fig:HP} for each mass bin, and report the best-fit values (which we take to refer to the median sample values from here on) and error bars on the 16th and 84th percentiles for all the hyperparameters in Table \ref{HPtable}. We find that radius inflation, $\Delta R = \mu_R-C$, governed primarily by $A$, is highly dependent on mass. The highest degree of inflation is found for planets with masses $0.37 - 0.98\si{\Mj}$, and it decreases with mass for the $0.98 - 2.5\si{\Mj}$ and $2.5+\si{\Mj}$ regimes. From the best-fit values of our parameters, the radius dependence at $F > F_s$ is:

\begin{eqnarray}
 \Delta R = 
 \begin{cases}
  -0.33 \cdot (\log_{10}F - 6.09), & \frac{M}{\si{\Mj}} < 0.37\\
  \label{low_bin_eq}
  0.70 \cdot (\log_{10}F - 5.5), & 0.37 \leq \frac{M}{\si{\Mj}} < 0.98\\
  0.52 \cdot (\log_{10}F - 5.8), & 0.98 \leq \frac{M}{\si{\Mj}} < 2.50\\
  0.22 \cdot (\log_{10}F - 5.2), & 2.50 \leq \frac{M}{\si{\Mj}}
 \end{cases}
\end{eqnarray}

\begin{figure}
 \centering
 \includegraphics[scale=0.5]{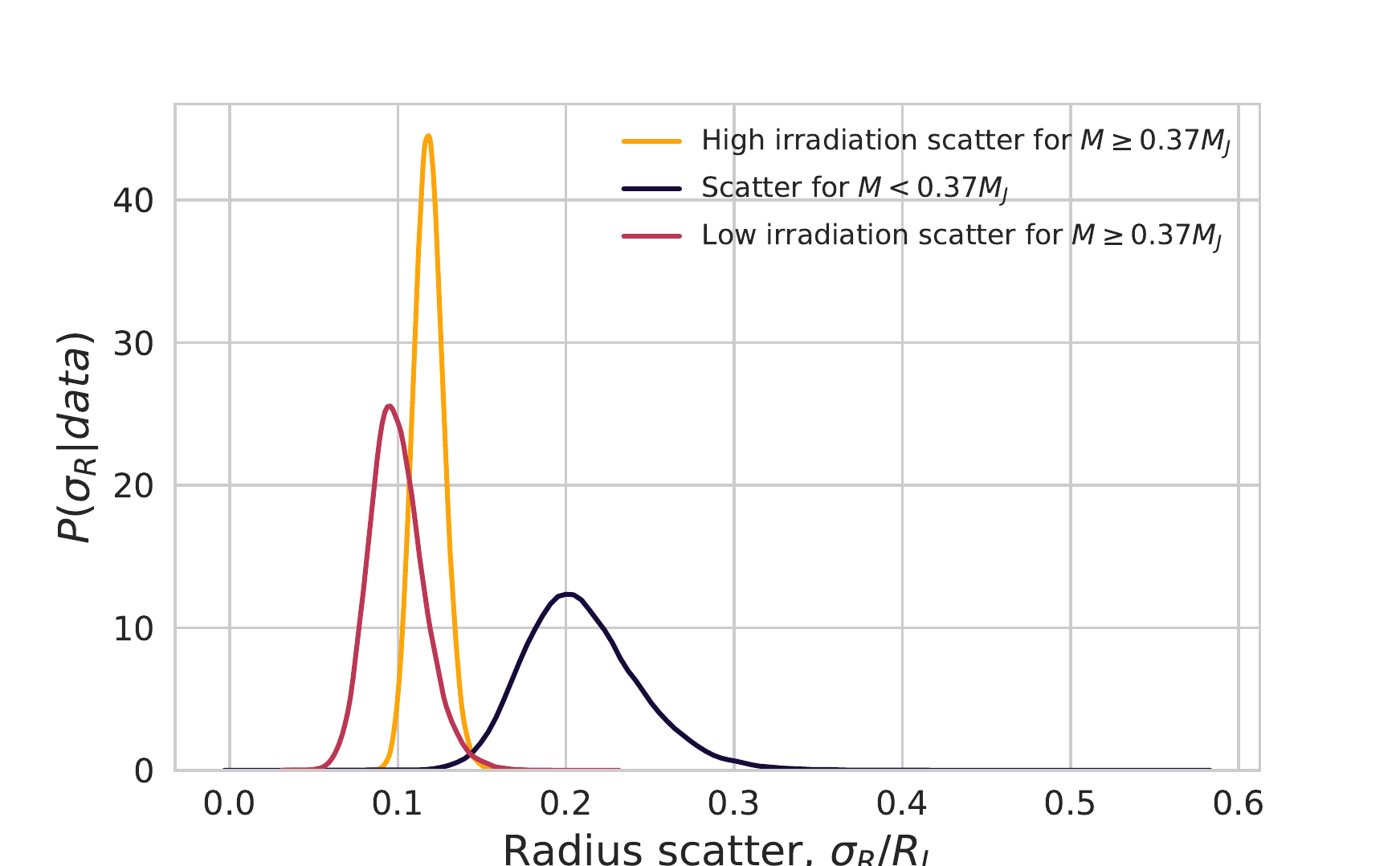}
 \caption{The posterior distributions for the intrinsic physical scatter for the 3 regimes considered (see Equation \ref{scatter_halfsplit}).}
 \label{scatter}
\end{figure}

\begin{figure}
 \centering
 \includegraphics[scale=0.5]{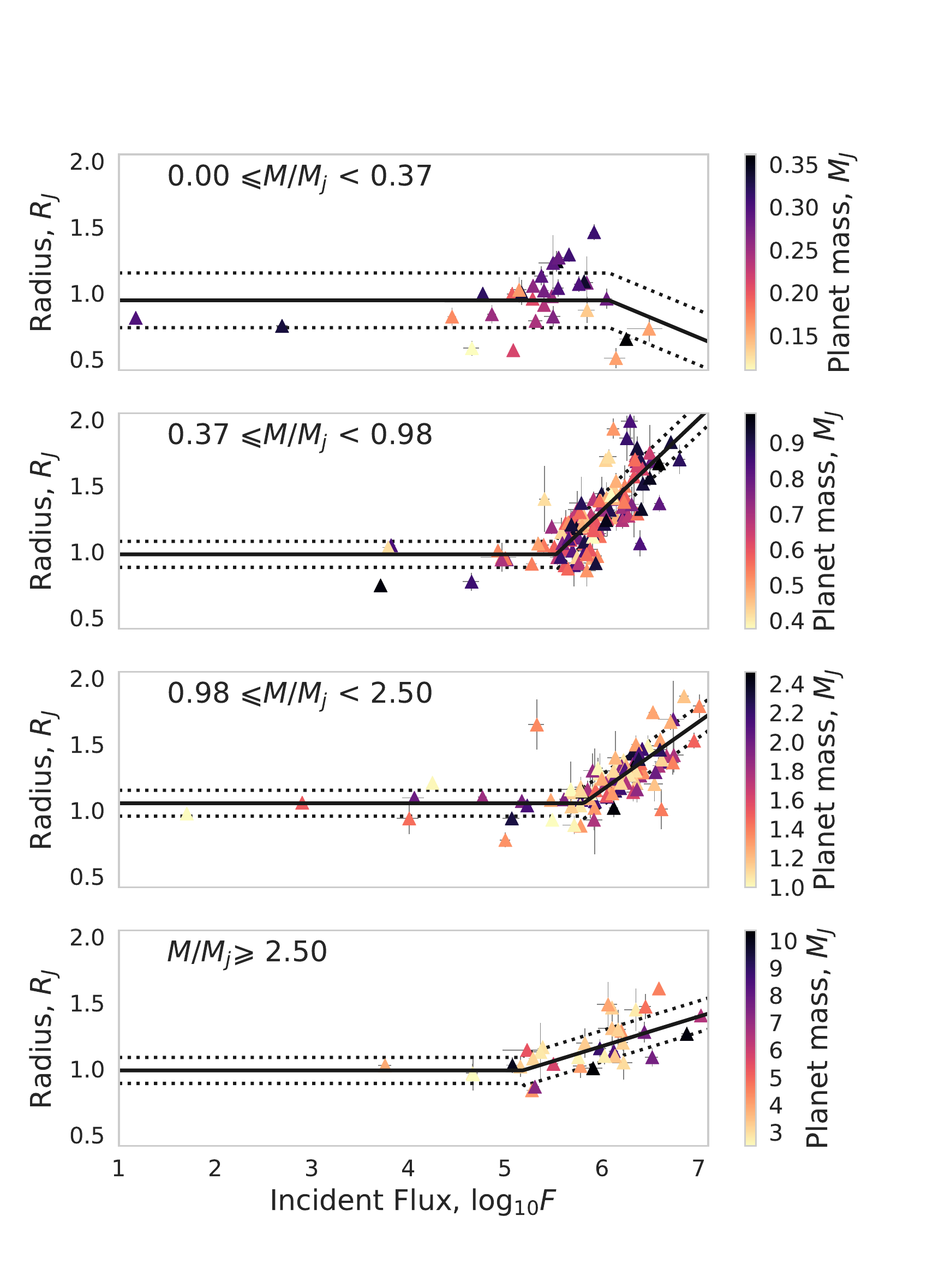}
 \caption{The relations of our model, plotted for parameters set at their best-fit values. The solid line is $\mu_R(F,M)$, and the dotted lines are $\mu_R(F,M)\pm\sigma_R(F,M)$ (in the best-fit model, 68\% of planets' true values should lie within). The points represent the data values of the planet parameters of our sample $(\widetilde{F}_i,\widetilde{M}_i,\widetilde{R}_i)$, with observational uncertainty error bars.  }
 \label{multirelation_data}
\end{figure}

\begin{figure}
 \centering
 \includegraphics[scale=0.5]{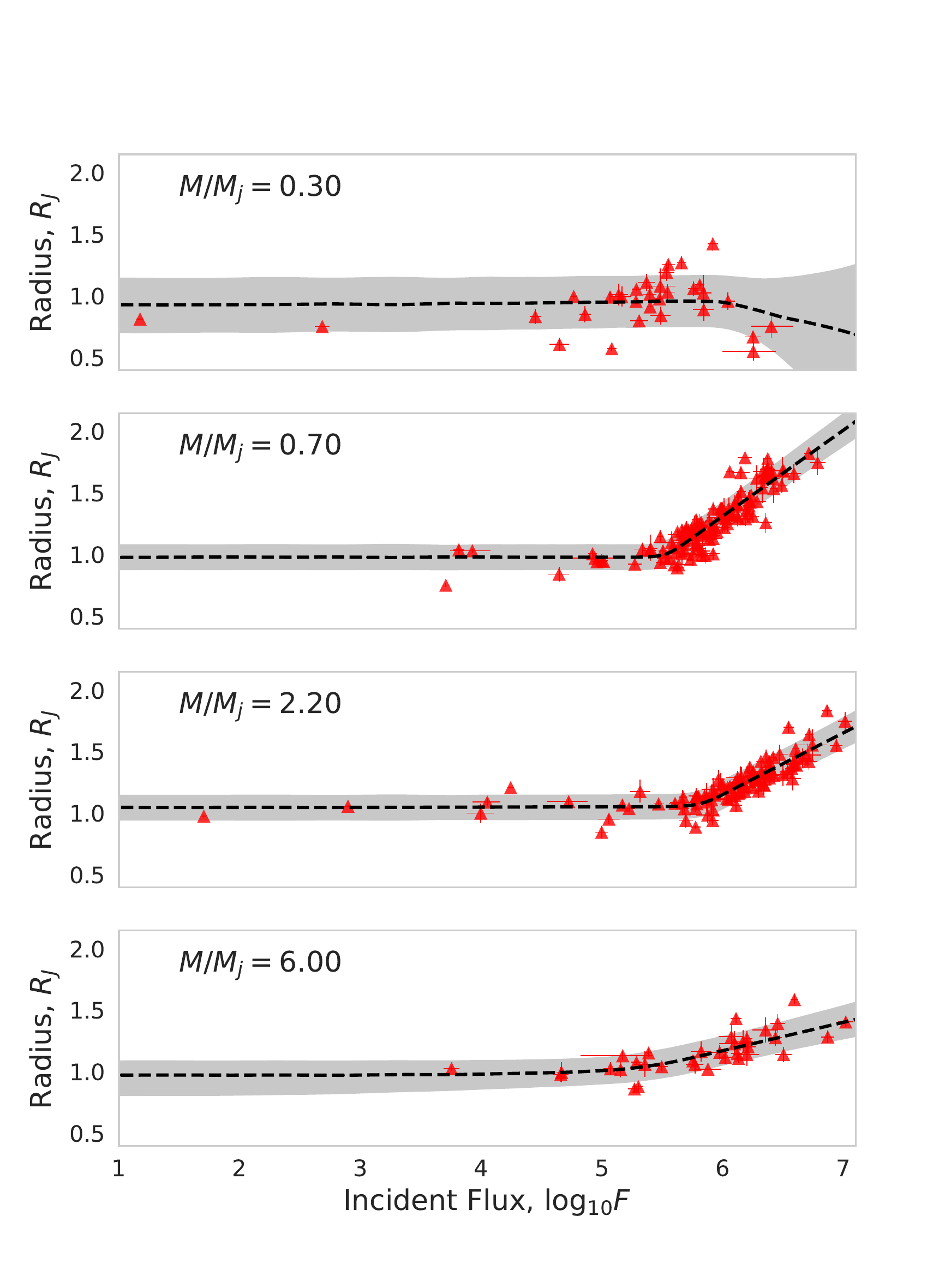}
 \caption{The marginalised posterior distribution of radii given flux, $p(R|F,M,\tilde{\boldsymbol{x}})$, plotted for different planet masses, see Equation \ref{true_ppf}.
 The shaded region is the central $68\%$ interval, within which $68\%$ of planet true radii should lie for a given incident flux and planet mass. It represents the $1\sigma$ points of $p(R|M,F,\tilde{\boldsymbol{x}})$; the central dashed line is the median and the dotted lines are the $95\%$ interval. The red points represent the posterior true values of the planet parameters of our sample, with posterior uncertainty error bars.}
 \label{multirelation_post}
\end{figure}

For the lowest mass bin, $M<0.37\si{\Mj}$, the posteriors are much broader and the scatter is much larger, shown in Fig. \ref{scatter}. In Fig. \ref{multirelation_data} and \ref{multirelation_post}, we see that the radius-flux distribution is more complicated than for the other mass bins, featuring a change in the behaviour of the radii at $\sim \si{10^6}{Wm^{-2}}$. A number of planets appear to not follow the trend of increasing radii with flux\footnote{The 5 planets are HD 149026 b \citep{butler2006_catalog,hd149026b_mass}, K2-39 b \citep{k239b}, Kepler-101 b \citep{kepler101b}, Kepler-41 b \citep{kepler41b_radius,butler2006_catalog} and WASP-126 b \citep{wasp126b}.}, and we also note a lack of inflated planets at fluxes higher than $\sim \si{10^6}{Wm^{-2}}$ in this mass range. Our chosen model, which assumes that the planets in a mass range behave uniformly with regards to their radius relationship with flux, doesn't seem to be a good fit for the least-massive planets in our sample. We caution the reader that due to its large uncertainties, the model fit below $0.37\si{\Mj}$ may not be reliable. However it is still informative, as it places a lower bound on the range of masses over which our model works. For a further discussion see Sect. \ref{low_mass}.

\begin{center}
\begin{table}
\caption{ The best fit values for the hyperparameters, with error bars for the 16th and 84th percentiles. The mass boundaries were $0.37\pm0.02$, $0.98_{-0.05}^{+0.04}$ and $2.5$, with the latter fixed at that value.}
\centering{}%
\begin{tabular}{l  c c c }
\hline 
Mass Range & $A$ & $C$ & $log_{10} F_{s}$\tabularnewline
\hline 
$0.10 - 0.37$ & $-0.33_{-0.91}^{+0.51}$ & $0.95_{-0.08}^{+0.05}$ & $6.1_{-1.6}^{+0.7}$\tabularnewline
$0.37 - 0.98$ & $0.70_{-0.06}^{+0.07}$ & $0.98\pm0.04$ & $5.52_{-0.09}^{0.07}$\tabularnewline
$0.98 - 2.50$ & $0.52_{-0.07}^{+0.07}$ & $1.06\pm0.03$ & $5.82_{-0.11}^{+0.09}$\tabularnewline
$2.50 +$ & $0.22_{-0.06}^{+0.08}$ & $0.99_{-0.14}^{+0.05}$ & $5.2_{-1.1}^{+0.3}$\tabularnewline
\hline
 & $\sigma_{R,0}$ & $\sigma_{R,1}$ & $\sigma_{R,3}$ \tabularnewline
\hline
 & $0.21_{-0.03}^{0.04}$ & $0.10_{-0.01}^{0.02}$ & $0.12\pm0.01$\tabularnewline
\hline
\end{tabular} \\
\label{HPtable}
\end{table}
\par\end{center}

We find that the mean low irradiation radius $C$ is highest for the class of planets more massive than Jupiter ($0.98 - 2.50\si{\Mj}$), and decreases slightly for both heavier and lighter planets, in line with what is expected from model predictions \citep{fortney2007,baraffe2008}. However it would also be affected by any trends in the heavy element masses of the planets \citep[see][]{miller2011,thorngren2016}.
 
The inflation threshold is only well constrained in $0.37-2.5\si{\Mj}$, and the inflation mechanism requires higher fluxes to be activated when mass is increased, as $F_s$ is about 1.3 times greater for the $0.98 - 2.50\si{\Mj}$ than for the $0.37 - 0.98\si{\Mj}$ gas giants. For $2.5+ \si{\Mj}$ the distribution of $F_s$ provides a poor constraint, with an extended tail at low incident fluxes in Figure \ref{fig:cp3}. This is likely due to the lack of heavy planets at low irradiation in our sample. Similarly for $0.1-0.37 \si{\Mj}$ there is also a poor constraint, with a long tail in Figure \ref{fig:cp0}, likely as a result of the poor fit of our model in this mass range.

Figure \ref{scatter} shows the physical scatter posteriors, $\boldsymbol{\sigma}_R$. Low mass planets ($M<0.37\si{\Mj}$) have by far the largest scatter, at $~0.21\si{\Rj}$. For masses above $0.37\si{\Mj}$, we find the scatter is slightly higher in the inflated planets above the flux threshold than for weakly irradiated planets ($0.12\si{\Rj}$ vs $0.10\si{\Rj}$). Most of the proposed mechanisms for inflation would introduce their own latent parameters which will control the degree of inflation, e.g. opacity, tidal heating rate and Ohmic heating depth. Variations in these parameters may add to the scatter already present from heavy element content \cite[see e.g][]{heng2012}.


\subsection{Marginalised posterior distribution}
\label{ppf}

The functional form of the FMR relations, taking the best-fit parameter values, is shown in Fig. \ref{multirelation_data}, with scatter limits as dotted lines. The points represent the raw observational data-set $\tilde{\boldsymbol{x}} = \{\widetilde{F}_i,\widetilde{M}_i,\widetilde{R}_i\}_{i=1} ^N$ and its observational uncertainties.

Figure \ref{multirelation_post} represents a more complete picture of our posterior FMR relation. The points are the posterior true parameters and their uncertainties, which can be seen to be different from the observed values. The shaded regions show the posterior predictive distribution for the true radii, and represent the parameter space within which we expect the central 68\% of planets to be contained for a given incident flux and planet mass. It displays the $1\sigma$ contour lines of the distribution $p(R|M_j,F,\tilde{\boldsymbol{x}})$ as a function of $F$ for specific masses\footnote{The masses of each plot are fixed and chosen to be in the middle of each mass bin (away from the boundaries). Near the boundaries, the scatter would be larger (and the shaded region wider) as it transitions from one bin to the next. This is because marginalising over the hyperparameters, including the mass bin boundaries, means that a particular mass could lie in either of the two adjacent mass bins with some probability. In essence, the $R|F$ relations plotted above vary smoothly with mass (except at the $2.5\si{\Mj}$ boundary), though only cross sections are shown in Fig. \ref{multirelation_post}.} 
 and is marginalised over the posterior distribution of the hyperparameters:

\begin{equation}
p(R|M,F,\tilde{\boldsymbol{x}}) = \int p(R|M,F,\boldsymbol{\alpha})p(\boldsymbol{\alpha}|\tilde{\boldsymbol{x}})d\boldsymbol{\alpha}
\label{true_ppf}
\end{equation} 

where $p(\boldsymbol{\alpha}|\tilde{\boldsymbol{x}})$ is the posterior distribution of our hyperparameters. The dashed line represents the median point of the distribution in Equation \ref{true_ppf}, and the dotted line is the central 95\% coverage interval. In the Bayesian sense, Equation \ref{true_ppf} represents the most accurate statement of our knowledge of the FMR relation, having inferred the posterior distributions of all our model parameters using the available data, as it includes the uncertainties we have in those parameters. 
However, the best-fit distribution $p(R|M,F,\hat{\boldsymbol{\alpha}})$, with $\hat{\boldsymbol{\alpha}}$ from Table \ref{HPtable} is often a good enough approximation.

The $\widetilde{R}_i$ values in Fig. \ref{multirelation_data} are more scattered than the true $R_i$ in Fig. \ref{multirelation_post}. This is expected (see in Sect. \ref{hbm}), since the scatter from observational noise is removed from the posterior distribution for $R_i$. While the form of model we choose will affect these posteriors, they will tend to always shrink towards a mean line. We justify this result in the model checking in Sect. \ref{check}. The posterior distribution should be seen as the distribution of "true" planet parameters most likely to reproduce the data when observational noise is applied to them.



\section{Model checking}
\label{check}

\begin{figure}
 \centering
 \includegraphics[scale=0.5]{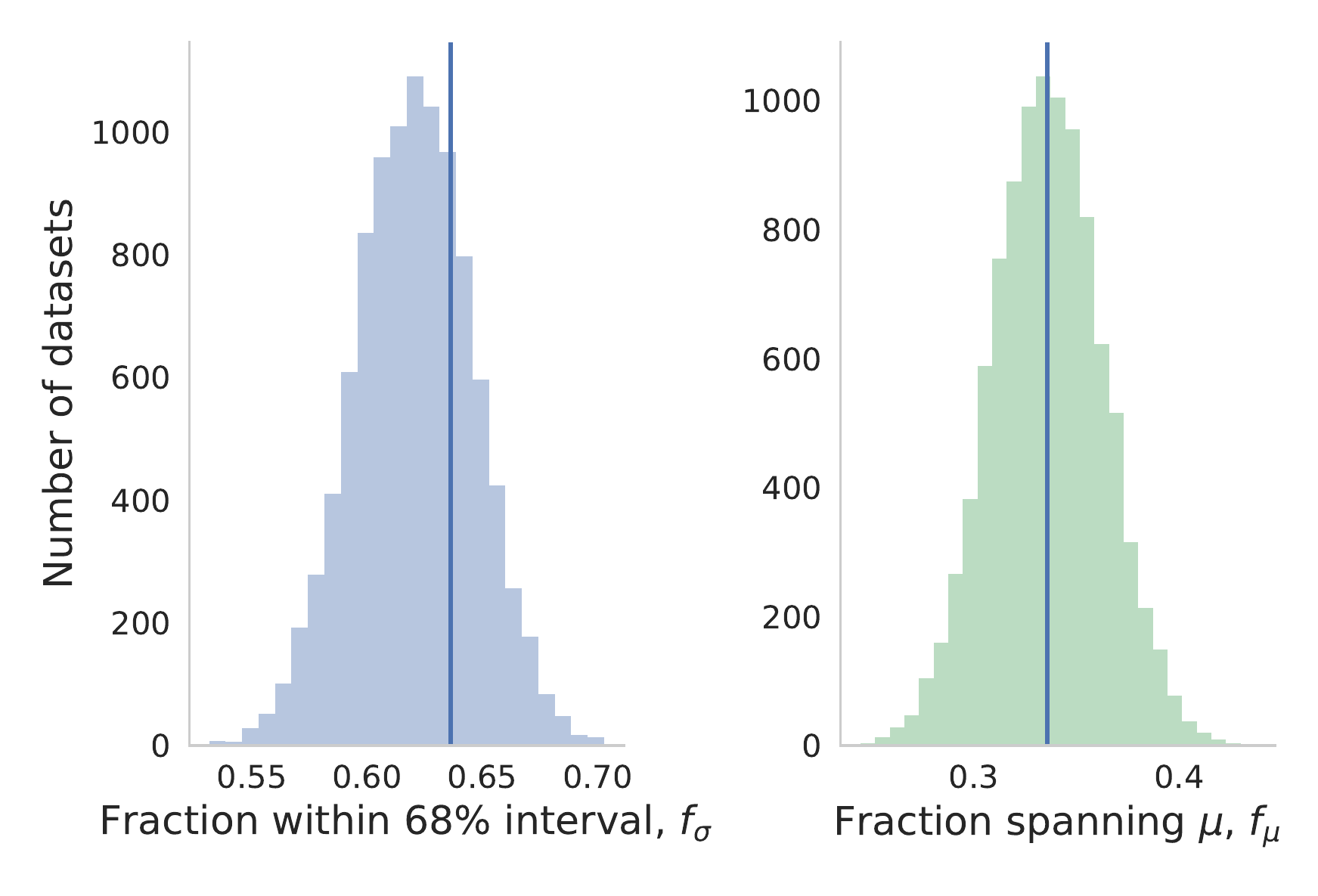}
 \caption{The distribution of $f_{1\sigma}$ (left) and $f_{\mu}$ (right) for 10,000 generated datasets from the second level of our HBM (the true posteriors, see equation \ref{level_two_eq}), with comparison to the values for the real dataset represented by the vertical line. The real data falls within the 73rd and 50th percentiles of the generated distributions for $f_{1\sigma}$ and $f_{\mu}$ respectively.}
 \label{level_two}
\end{figure}

\begin{figure}
 \centering
 \includegraphics[scale=0.5]{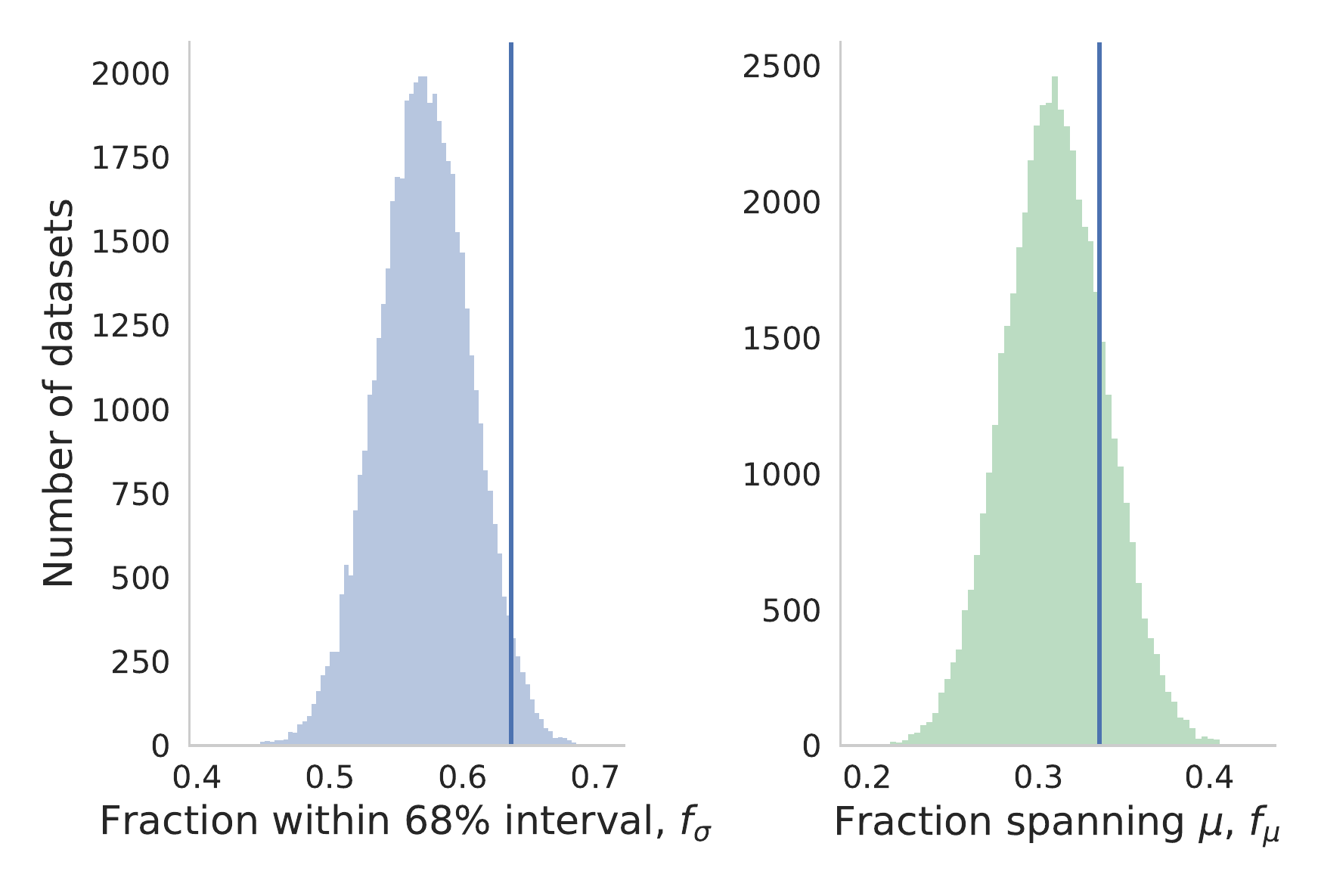}
 \caption{The distribution of $f_{1\sigma}$ (left) and $f_{\mu}$ (right) for 50,000 generated datasets from the third level of our HBM (the hyperparameters, see equations \ref{level_three_eq} and \ref{level_three_obs}), with comparison to the values for the real dataset represented by the vertical line. The real data falls within the 97th and 82rd percentiles of the generated distributions for $f_{1\sigma}$ and $f_{\mu}$ respectively.}
 \label{level_three}
\end{figure}

We perform detailed tests to justify our model and check its consistency with the data, by using a posterior-predictive test. Our aim is to test the ability of our model to reproduce data-sets ``similar'' to the observed data set, and to quantify the discrepancy. Due to the similar nature of our investigation, we follow the procedure of \cite{wolfgang2016}, to which we refer the reader for a more in-depth explanation. For a detailed overview of hierarchical model checking, see also \cite{bayarri2007}.\par 

We calculate the posterior predictive distribution of model $\mathcal{M}$; defined as the probability of observing a new planet with parameters $x_{new} = (\widetilde{R}_{new}, \widetilde{M}_{new}, \widetilde{F}_{new})$, given our original observed data set:

\begin{eqnarray}
 \label{posterior_p_fit}
p(x_{new}|\tilde{\boldsymbol{x}},\mathcal{M}) & = & \int p(x_{new}|\boldsymbol{\theta},\mathcal{M})p(\boldsymbol{\theta}|\hat{\boldsymbol{x}},\mathcal{M})d\boldsymbol{\theta}\\
 & = & \mathbb{E}_{\boldsymbol{\theta}|\hat{\boldsymbol{x}},\mathcal{M}} [p(x_{new}|\boldsymbol{\theta},\mathcal{M})]
 \label{expect_ppf}
\end{eqnarray}

where $\boldsymbol{\theta}$ refers to the combined model parameters ($\boldsymbol{R}$, $\boldsymbol{M}$, $\boldsymbol{F}$, $\boldsymbol{\alpha}$). Equation \ref{expect_ppf} is to be read as the expectation value of $p(x_{new}|\boldsymbol{\theta},\mathcal{M})$, for values of $\boldsymbol{\theta}$ drawn from their posterior distribution, $p(\boldsymbol{\theta}|x_{new},\mathcal{M})$. It follows from Equation \ref{posterior_p_fit} by the definition of the expectation value (mean in this case), and is of course trivial to calculate with the samples produced by our MCMC. From here on, we will drop the $\mathcal{M}$ notation and treat it as implicit.\par 

To perform the posterior-predictive test, we draw points from the posterior predictive distribution in Equation \ref{posterior_p_fit} to create a mock data set, $\tilde{\boldsymbol{x}}_{new}$, of the same size N as our planet sample. By drawing thousands of such mock datasets, we can compare them statistically to our observed dataset. \par

The comparison in this case involves quantifying certain aspects of the distributions in some derived "statistics". We use the same two statistics as \cite{wolfgang2016}: $f_{1\sigma}$, the fraction of a given dataset's simulated radii that fall within the 68\% coverage interval (the shaded region in Fig. \ref{multirelation_post}, described in Sect. \ref{ppf}), and $f_{\mu}$, the fraction of data points which have radius $1\sigma$ error bars that cross the median of the posterior distribution (the dashed line in Fig. \ref{multirelation_post}). While we are using the posterior true distributions to aid in performing this comparison, it is important to remember that what we are comparing is the "observed" data, i.e $\widetilde{R}$, produced from the true values (e.g, $R$), based on observational uncertainties.\par

The two statistics test two key attributes of our model. $f_{1\sigma}$ is a measure of the scatter of data produced by our model (controlled mainly by $\sigma_R$), while $f_{\mu}$ is a measure of how centered the data is about the median line. $f_{\mu}$ is thus a proxy for the shape of the data set distribution, and checks whether the normal distribution chosen in equation \ref{R} is justified. As an example, if the scatter of our model proved to be accurate (in reproducing $f_{1\sigma}$), but $f_{\mu}$ was too small, it could suggest that we should have used a distribution with heavier tails and less clustering near the mean.\par

We must also address how to draw a posterior predictive dataset, $\boldsymbol{x}_{new}$, from Equation \ref{posterior_p_fit}, where we also follow the treatment of \cite{wolfgang2016}. There are two methods, each tied to one level of our hierarchical Bayesian model, where the second level represents the true values ($\boldsymbol{R}$,$\boldsymbol{F}$,$\boldsymbol{M}$) and the third level represents the hyperparameters $\boldsymbol{\alpha}$ (see Fig. \ref{bayesnetwork}). \par

For the first method (drawing from the second level), we draw each mock dataset from the posterior true value distributions of the real planet sample. Thus for the $i^{th}$ planet of our mock dataset, we draw $x_{new,i}$ from:

\begin{eqnarray}
x_{new,i} & \sim & \int p(x_{new,i}|R_i,M_i,F_i)p(R_i,M_i,F_i|\boldsymbol{x})dR_idM_idF_i\\
& \sim & \mathbb{E}_{R_i,M_i,F_i|\hat{\boldsymbol{x}}} [p(x_{new,i}|R_i,M_i,F_i)]
\label{level_two_eq}
\end{eqnarray} 

where 

\begin{equation}
p(x_{new,i}|R_i,M_i,F_i) = p(\widetilde{R}_{new,i}|R_i)p(\widetilde{F}_{new,i}|F_i)p(\widetilde{M}_{new,i}|M_i)
\end{equation}

represents the published observational uncertainty distributions of that particular planet. Performing the test on the datasets produced above checks the posterior distributions of our true parameters for the real planet population (i.e. the points in Fig. \ref{multirelation_post}).\par

The second method generates an entirely new population of hypothetical planet true values from the marginalised model:

\begin{eqnarray} 
R_i & \sim & \int p(R_i|M_i,F_i,\boldsymbol{\alpha},\boldsymbol{x})p(\boldsymbol{\alpha}|\boldsymbol{x})d\boldsymbol{\alpha}\\
& \sim & \mathbb{E}_{\boldsymbol{\alpha}|\hat{\boldsymbol{x}}} [p(R|M,F,\boldsymbol{\alpha})]
 \label{level_three_eq}
\end{eqnarray}

and from that hypothetical planet sample, we generate a new dataset as in the first method: 

\begin{equation}
x_{new,i} \sim \mathbb{E}_{R_i,M_i,F_i|\hat{\boldsymbol{x}}} [p(x_{new,i}|R_i,M_i,F_i)]
\label{level_three_obs}
\end{equation}

In this case, we draw the experimental uncertainties from distributions consistent with our data, conserving the trends between the parameter uncertainties and parameter values.\footnote{$\sigma_{M,obs}$ and $\sigma_{F,obs}$ are strongly correlated with $M$ and $F$ respectively. We thus fit a linear regressive model (including a dispersion term) for $\log \sigma_{M,obs} \sim N(f_1(\log M),\sigma)$ and $\log \sigma_{F,obs} \sim N(f_2(\log F),\sigma)$, and we draw $\sigma_{M,obs}$ and $\sigma_{F,obs}$ from the resulting highest-likelihood distribution. We also find that $\sigma_{R,obs}$ is correlated with $F$. To assign $\sigma_{R,obs}$, we split the range of fluxes into four bins, and draw $\sigma_{R,obs}$ from a random planet that is in the same $F$ bin.} 
Our hyperparameters do not parametrise the distributions of $F$ and $M$ for our population, thus we must first draw $M$ and $F$ from the same distributions as our population to use them in Equations \ref{level_three_eq}  and \ref{level_three_obs}. Performing the model checking statistics on the hypothetical datasets produced by this method is thus a test of our posteriors for the third level of our HBM, i.e the hyperparameters.

We calculate $f_{1\sigma}$ and $f_{\mu}$ for a large number of datasets produced by both the above methods, and also for the single real dataset. If our model is consistent with the data, we expect that the statistics calculated from the mock datasets and from the real data agree with each other. We define "agreement" as meaning that the statistic calculated for the real dataset is within the $1\sigma$ interval of the statistics calculated from the mock datasets. In other words, the real dataset must have the same properties as a "typical" mock dataset.



Figure \ref{level_two} shows the results of our model checking for the second level of our HBM, and Fig. \ref{level_three} shows the results for the third level. We measure model-data consistency by computing the percentile of the generated mock datasets that the real datasets falls into (measured by the statistic of choice).\par 


We find that the observed $f_{\mu}$ fall within the typical values produced by our model for datasets produced by both methods, coming in at the 50th and 82nd percentile for the second and third levels respectively. The observed $f_{\sigma}$ falls in the 73rd percentile of data generated from the second level of our HBM, indicating that the spread of our posterior true values $(R_i,M_i,F_i)$ is consistent with the data, and justifying the shrinking effect seen in the points of Fig. \ref{multirelation_post}. For the data generated by the third level of our HBM,  $f_{\mu}$ falls in the 82nd percentile, while $f_{\sigma}$ falls in the 97th percentile. In other words, only $\sim 6\%$ of the generated mock data were more extreme than our observed data. The result suggests that the form of our model  may have room for improvement, as we overestimate the scatter of the data. However, it does not negatively affect our conclusions in Sect. \ref{inflation}. \par 



\section{Discussion}
\label{discussion}

\subsection{The flux-mass-radius relation}
\label{trend}

Our results show a strong trend of decreasing radius inflation with mass, with the most inflated planets being in the mass range $0.37-0.98\si{\Mj}$ (see Fig. \ref{preplot}). This effect is also expected from models without additional inflation mechanisms, with mass increases past a few Jupiter masses leading to decreasing radius response to incident flux \citep{fortney2007,baraffe2008}. Below the inflation threshold, we also find that planet mass has a small effect, with the largest cold giant radii having masses of $0.98-2.50\si{\Mj}$.

Findings by \cite{thorngren2016} show that the heavy element content of non-inflated gas giants is strongly and positively correlated with planet mass. Unless there is a different formation mechanism behind the hottest Jupiters, or in cases of significant evaporation, this trend should apply in our sample as well. This would lead to a further reduction in the radii as planet mass increases, and would agree with the observed trends for inflated hot Jupiters. As more detailed predictions from models with inflation become available in the future (with predictions given for wide ranges of parameters such as in \cite{fortney2007}), we will similarly be able to probe the core mass distribution of inflated hot Jupiters.\par

\begin{figure}
 \centering
 \includegraphics[scale=0.5]{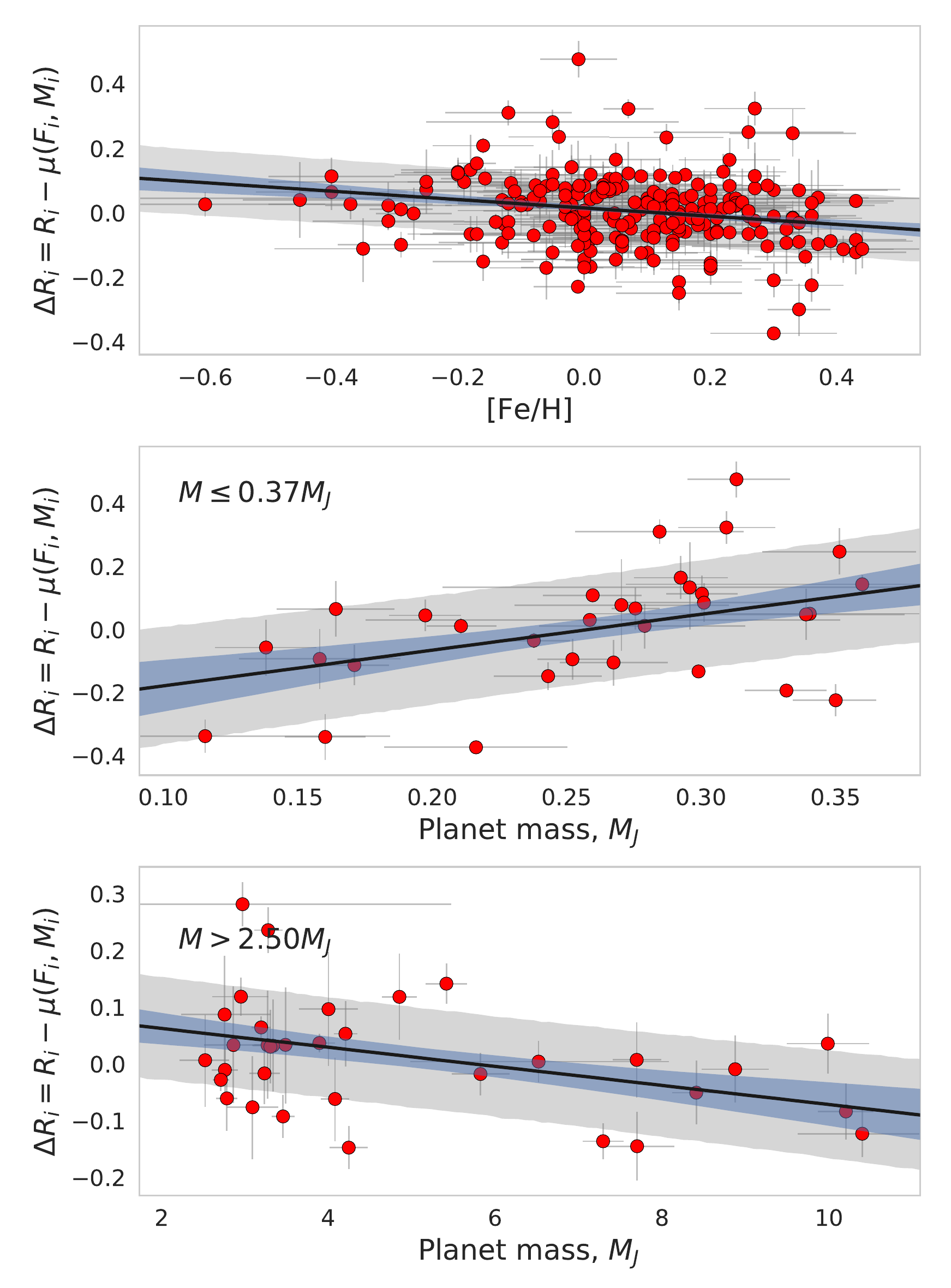}
 \caption{Radius residuals, $\Delta R_i = R_i - \mu(F_i,M_i)$, plotted against host star metallicity (top), and planet mass for the lowest mass bin (middle), and the highest mass bin (bottom). The shaded grey region is the marginalised 68\% coverage interval of our linear regression, and the shaded blue region represents the $1\sigma$ limits of  the mean line.}
 \label{mass_res}
 \label{met_res}
 \label{residuals}
\end{figure}

The incidence of using binning in the mass parameter must also be examined. To do this we look at the radius residuals, $\Delta R_i = R_i - \mu_R (F_i,M_i)$, and their relation with mass. We fit a simple linear regression to the relationship, with an additional scatter parameter, in order to determine whether a mass-radius gradient of zero is excluded to more than $1\sigma$. We find that within the $0.10-0.37\si{\Mj}$ bin there is a strong positive correlation of radius with mass, and a negative correlation for the $2.5 +\si{\Mj}$ planets (see Fig. \ref{mass_res}), and for both a gradient of zero is more than $1\sigma$ from the mean of the distribution. This additional dependence with mass may be the reason for the large scatter in the lowest mass bin, $M < 0.37$. We find no evidence of a mass-radius correlation for the middle two mass bins. Future models could take this into account.\par

Extending this analysis to other parameters, we also find a weak negative correlation of radius with metallicity for all mass bins (see Fig. \ref{met_res}), albeit with a very large scatter. This could be caused by a correlation between heavy element content and stellar metallicity \citep{guillot2006,miller2011}. However, we found no correlation of the radius with host star age. Our results agree broadly with \cite{enoch2012}.

We also perform a short test to determine the first-order bias introduced by averaging our observational error bars where they are asymmetric (see Section \ref{data}). The asymmetry in the uncertainties is most significant for the radii, where it is biased towards larger upper error bars. To determine the bias, we fit a simple 3-level hierarchical model to the residual observed radii, $\Delta \widetilde{R}_i = \widetilde{R}_i - \mu_R (F_i,M_i)$, treating the population as being drawn from a normal distribution with mean $\mu$ and variance $\sigma^2$, and having the same measurement uncertainties as the data. We then take two cases, one where we average the upper and low uncertainties as before, and one where we allow them to vary. For the latter, we use a discontinuous split normal distribution:

\begin{equation}
p(\Delta \widetilde{R}_i|\Delta R_i) = 
\begin{cases}
  Normal(\Delta \widetilde{R}_i, \Delta R_i, \sigma_{l}^2), &  \Delta R_i < \Delta \widetilde{R}_i\\
  Normal(\Delta \widetilde{R}_i, \Delta R_i, \sigma_{u}^2), &  \Delta R_i \geq \Delta \widetilde{R}_i
 \end{cases}
\end{equation}

to simulate asymmetric uncertainty distributions with upper and lower error bars $\sigma_u$ and $\sigma_l$ respectively. Comparing the results between the two cases, we find that the difference in our posteriors for $\sigma$ and $\mu$ are $0.05\%$ and $0.1\%$ respectively. This is well within the error bars on the posteriors in this simple HBM, and within the errors bars we present for our hyperparameters in Sect. \ref{results}, which are all on the order of $1-10\%$.



\subsection{Low-mass inflation cutoff}
\label{low_mass}


\begin{figure}
 \centering
 \includegraphics[scale=0.5]{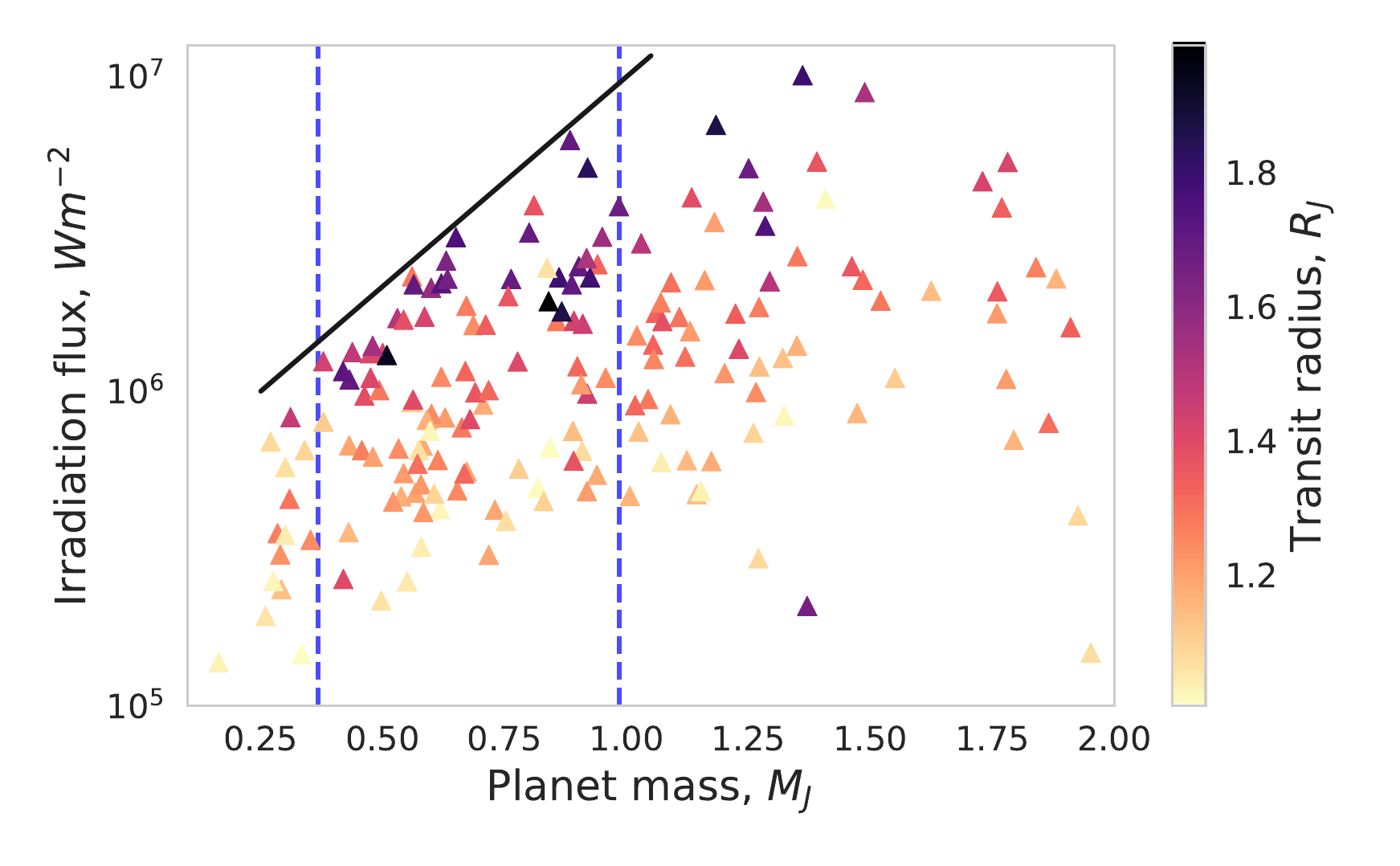}
 \caption{The mass-flux distribution for all gas giants that have a radius greater than $1\si{\Rj}$. The blue dashed lines show our best-fit mass boundaries. As the mass decreases below $1\si{\Mj}$, the maximum incident flux at which we find inflated planets also decreases (black line).}
 \label{fmrelation}
\end{figure}

A key finding we present is the inference of a boundary at $\sim 0.37\si{\Mj}$, above which we see the most inflated gas giants, with a clear positive dependence on incident flux. Below we find a lack of highly inflated radii with $R>1.5\si{\Rj}$, and a more complicated trend (see Figs. \ref{multirelation_data} and \ref{multirelation_post}). 


From visual inspection of the data in the lowest mass bin, we see the beginning of a trend of inflating radii as a function of flux $\sim \si{10^5}{Wm^{-2}}$, particularly for planets closer to $0.37\si{\Mj}$. However this stops at $\sim \si{10^6}{Wm^{-2}}$, and there are no inflated planets beyond this threshold. Instead we see a reversal, with decreasing radii and denser planets, although we have a small sample size in that regime (5 planets).  Giants at the top end of the mass range, which would be otherwise inflated, also show decreasing radii (e.g, HD 149026 b has a mass of $0.37 \pm 0.03\si{\Mj}$ and evolutionary models find it to be very dense, with a predicted core of $80$-$110\si{\Me}$, \cite{burrows2007}). Previous studies about hot Jupiter inflation \citep[e.g][]{laughlin2011,enoch2012} generally use manually chosen minimum mass cutoffs for their samples. However, we recommend that future studies focus on planets heavier than $\sim 0.37\si{\Mj}$, to avoid the samples being contaminated with planets where other processes may start to play a role alongside inflation.

In Fig. \ref{fmrelation} we plot the incident flux of all the planets in our sample that have radii greater than $1\si{\Rj}$ (and are thus potentially inflated), and find that there is a maximum incident flux that depends on the planet mass, and decreases as we decrease the mass. 


To explain the reversal of the inflation trend, and the lack of such inflated low mass giants at $F > 10^6 \si{Wm^{-2}}$, an obvious cause could be bulk evaporation or Roche-lobe overflow leading to significantly denser planets (if the outer gaseous layers are stripped). This could also lead to such evaporated giants leaving remnants too low in mass to enter our sample ($M<0.1\si{\Mj}$). Roche-lobe overflow and asymmetric evaporation, such as described in \cite{gu2003, baraffe2005} could also stop planetary migration beyond a certain point, or even lead to outward migration, which may explain the paucity of planets in this mass range closer to their star (relative to the other mass ranges).

We must also consider the observational biases which may be present. There are few data points available in this regime, however our conclusion relies not only on the trend of over-dense planets closer to the star, but also on the clear lack of $R > 1.0\si{\Rj}$ gas giants at incident fluxes greater than $\sim 10^6\si{Wm^{-2}}$. Below this threshold, there are numerous such large giants, and we would expect that as orbital distance is decreased and incident flux increased, the probability of detecting similar objects should increase. Similarly, larger radii should also be easier to detect, thus if observational surveys are able to find planets with $R < 1.0\si{\Rj}$ in the mass ranges that we examine, we would also expect them to be able to find more inflated planets (if they exist). Thus it is unlikely that our conclusions are impacted by an observational bias.\par

\subsection{Lack of non-inflated hot Jupiters}
\label{inflation}

Of the proposed mechanisms for radius inflation, many can only work under specific circumstances. For example, tidal dissipation requires non-zero eccentricity, unless the eccentricity can be excited \citep{arras2010}, and Ohmic heating requires heat deposition deep enough near the radiative-convective boundary, see e.g, \cite{heng2012,perna2012,ginzburg2016}. Thus, not only may we expect some planets to not be inflated, but the varying degrees of inflation caused by each mechanism should leave an imprint on the distribution of radii and its scatter. Here, as an example, we make a simple comparison of our posterior FMR distribution with planetary models that don't include inflation mechanisms \citep[i.e][]{fortney2007} to determine what fraction of radii could be explained without resorting to inflation. \par

To compare our observation-based results with theory, we must not only account for the mass and incident flux, but also for planet age, and the unknown core mass. We assume that planet ages are distributed uniformly in the range $0.316-8.0\si{Gyr}$ (in rough agreement with the age distribution of our planet sample). The core mass, or heavy element fraction, is not an observable quantity. However \cite{thorngren2016} showed that there exists a strong correlation between core mass and planet mass at low insolation, and inferred the relation through a regression fit. For the purposes of this work, we assume that the inferred core mass ($M_C$) to planet mass ($M$) relation of \cite{thorngren2016} holds equally well for highly irradiated planets. Their relation is denoted as a probability distribution $p_T(M_C|M,\hat{\boldsymbol{\theta}}_T)$
\footnote{The fitted model in \cite{thorngren2016} has the form $M_C/\si{\Me} = k \times 57.9 \times M^{0.61}$, where the multiplicative scatter, $k$ is drawn from $\log_{10}k \sim N(\mu=0,\sigma=\log_{10}1.83)$.}
, for the highest-likelihood point of their model parameters $\hat{\boldsymbol{\theta}}_T$.\par 

\cite{fortney2007} give a detailed table of predictions of planet radii as a function of mass, incident flux (orbital distance), core mass and planet age. However using the \cite{thorngren2016} relation, planets heavier than $\sim 1\si{\Mj}$ can have core masses greater than the domain of predictions in \cite{fortney2007}. Thus, we can only carry out the following analysis for the lowest inflated mass bin, $0.37-0.98\si{\Mj}$. There is also a systematic error in this comparison, as the models in \cite{fortney2007} consider the heavy elements to be contained fully in a rock-ice core, while \cite{thorngren2016} have the heavy elements also distributed in the envelope. Models with all the heavy elements contained in the core tend to produce larger radii than models with heavy elements mixed throughout the envelope, so our model radii are overestimated \citep[see][]{baraffe2008,thorngren2016}. \par 

\begin{figure}
 \centering
 \includegraphics[scale=0.5]{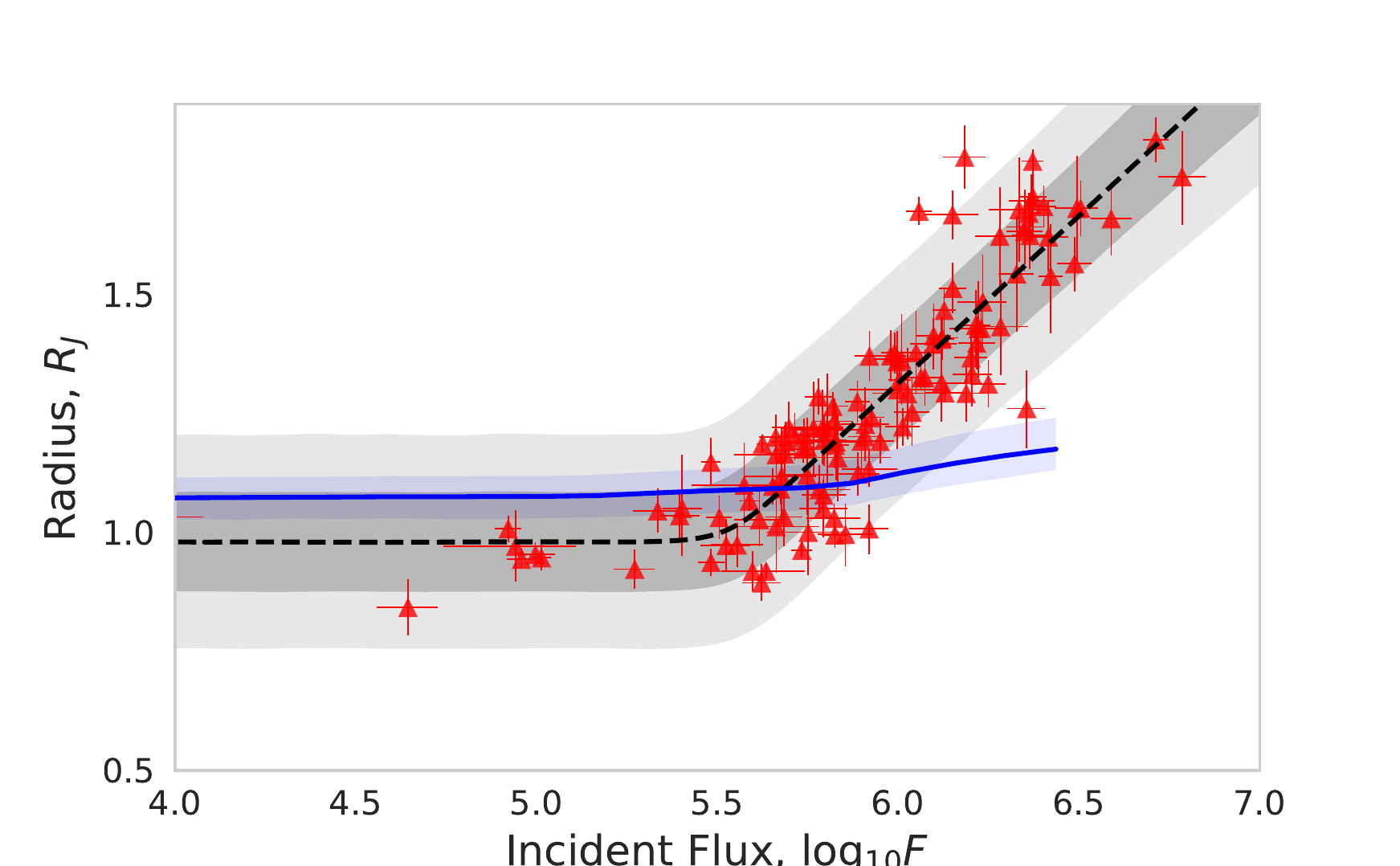}
 \caption{Our marginalized posterior radius-flux distribution for $0.37 \leq \frac{M}{\si{\Mj}} < 0.98$, with the central 68\% interval shaded in dark grey, and the 95\% interval in light grey. Plotted against a flux-radius relation $p_M(R|F)$ predicted by a non-inflationary planet model, with the central 68\% interval shaded in blue.}
 \label{comparison_fortney1}
\end{figure}

We produce a distribution of model radii as a function of incident flux, $p_M(R|F)$, fully marginalised over the relevant distributions of core mass and planet age ($t$) and marginalised over the width of the mass bin, $m_i-m_{i+1}$:


\begin{equation}
p_M(R|F) = \int \delta (R - f(M, M_C, t, F)) p_T(M_C|M,\hat{\boldsymbol{\theta}}_T)U(t)dtdMdM_C
\end{equation}

where $f(M, M_C, t, F)$ refers to the model predictions of \cite{fortney2007}, which are deterministic, $\delta$ is the delta-distribution, and $U(t)$ is the uniform distribution for the planet age. Thus we calculate the model prediction for the distribution of radius as a function of incident flux, similar to our posterior FMR distribution. We compare the model prediction against our observation-driven posterior FMR distribution in Fig. \ref{comparison_fortney1}.\par 

This is intended as a rough and qualitative exploration of how we could compare marginalised theoretical models to uncertainty decoupled data. We see that below the flux threshold, our theory-predicted $p_M(R|F)$ overestimates the radius, but falls within the $1\sigma$ interval. This could be due to us underestimating the core masses or planet age, but may also be cause by the previously mentioned increase in radii when all the heavy elements are placed into a pure rock-ice core, compared to being distributed in an envelope.\par 

With that in mind, we look at the predictions above $\sim 10^6\si{Wm^{-2}}$. The posterior FMR relation and the model prediction drastically diverge, with negligible overlap beyond $F \sim 1.6 \times 10^{6} \si{Wm^{-2}}$. Only one real planet, HATS-9 b \citep{brahm2015}, has a posterior density that has significant overlap with $p_{M}(R|F)$, and we cannot rule out the fact that it may be an inflated planet with an unusually large core. Such over-dense planets have been found in the low-flux regime, e.g, see Kepler-539 b, with total mass $1.00\si{\Mj}$ and heavy element mass $0.49\si{\Mj}$ \citep{thorngren2016}. Furthermore, what little overlap exists is only significant above the mean line of $p_M(R|F)$ (the solid blue line). Below the mean line, where we would expect to find denser planets with large cores, there is negligible overlap, even with the 95\% interval of our HBM's FMR relation.\par 

While a more detailed study would be warranted, our results suggest that classical planets which do not exhibit radius inflation do not exist above fluxes of $\sim 10^6\si{Wm^{-2}}$ for this mass range. Even if some of the observed planets could be explained as being young non-inflated gas giants with light cores, if a population of such young low-density planets were to exist, we would also expect to find similar planets with medium-mass or heavy cores and at older ages (which would fall below the mean line of $p_M(R|F)$).\par 

Although we found a slight discrepancy between our hyperparameter-marginalised HBM posteriors and the data in Sect. \ref{check}, this discrepancy suggests that our HBM is overestimating the intrinsic scatter. With a lower scatter, the overlap between the observed data and the model prediction would be decreased. Although we found that we were overestimating the radii at low irradiation, increasing the input core masses to counter this would only serve to further drive $p_M(R|F)$ away from our FMR distribution at high fluxes.\par

If re-inflation of hot Jupiters is impossible \citep[see][]{wu2013, ginzburg2016}, this would mean that all of the current population of highly irradiated $0.37-0.98\si{\Mj}$ giants must have migrated to their present locations very early in their lifetimes. This could however be avoided with tidal re-inflation, e.g \cite{jermyn2017}, followed by delayed contraction due to Ohmic heating or another mechanism.\par 

Furthermore, the findings of \cite{heng2012} show that variations in atmospheric scattering and absorption should produce a large scatter in the degree of Ohmic dissipation. In certain cases (such as atmospheres with temperature inversions and high optical opacities, or planets with too weak a magnetic field) we may even expect no significant Ohmic dissipation at depth \citep{heng2012,perna2012}. This should be reflected in the radius distribution of hot Jupiters. Thus if Ohmic heating is the dominant inflation process, the lack of un-inflated hot Jupiters must be explained by either a lack of high optical opacities and temperature inversions in this temperature regime, or would require a different inflation mechanism for the cases where Ohmic heating might fail.\par


We also briefly consider the observational biases that may be present in our data. While there is a detection bias towards finding larger radii and masses that we can't fully incorporate into our model, since we use various sources of data for which we may not have well-documented completeness functions, we note that the un-inflated planets that we are comparing with have roughly the same radii as gas giants found below the inflation threshold. Thus surveys that can detect the giants below the flux threshold should detect closer-in gas giants of the same radii as well. Planets closer to their star should also have a higher chance of transiting, and thus being detected, so we should be biased towards finding planets in the high incident flux regime.


A more thorough treatment of the observational biases would be the next step, as well as a comparison with newer models that include inflation mechanisms. Within the framework of such a hierarchical model, we could also extract a more quantitative measure of the percentage of gas giants that are un-inflated using a mixture model \citep[see][]{mixture_models_2000,celeux2007}. Another approach, applied recently by \cite{thorngren2017}, is to infer the heating required without specifying an underlying mechanism. They find that the functional form of the heating rate $\epsilon (F)$ required to explain the radii most closely matches the profile predicted by Ohmic heating. Given our findings that all $0.37-0.98\si{\Mj}$ hot-Jupiters beyond $\sim 10^6\si{Wm^{-2}}$ show inflation, it would be informative to further explore the scatter of $\epsilon (F)$ at a particular flux within their framework.


\section{Conclusions}
\label{conclusion}

We have constructed a theory-independent hierarchical Bayesian model to constrain the probabilistic relation between the planet radius, mass and incident flux for 286 gas giants. We thus find the intrinsic scatter of the gas giant population, decoupled from the observational uncertainties in both the radius parameter, and the mass and incident flux parameters. The posterior distribution we have inferred may be used to test and constrain theoretical models for gas giant radius evolution, especially involving inflation mechanisms. Constraining the intrinsic scatter allows us to determine the real range of radii that need to be reproduced by theory at each regime of incident flux and mass. \par

Our key results are summarized below:

\begin{itemize}
\item We find that planetary mass plays a significant role in the degree of radius inflation, with the most inflated hot Jupiters coming from $0.37-0.98\si{\Mj}$ range, and that the response of radius to incident flux decreases as mass is increased. This could be caused by either a trend of increasing core masses with increasing total mass, similar to that found for gas giants below the inflation threshold \citep[see][]{thorngren2016}, or by the increase in surface gravity with mass, which would make them harder to inflate.
\item Below $0.37$ Jupiter masses, there is an abrupt lack of heavily inflated radii, and we find that radii begin to decrease with incident fluxes near $\sim 10^6\si{Wm^{-2}}$. Such a trend of decreasing radii is not present at higher masses. We also note that there is a cutoff point in incident flux beyond which hot Jupiters no longer  exist, and that for inflated planets below $1.0\si{\Mj}$ this cutoff point decreases with decreasing mass.
\item We use our inferred posterior distribution to show that there is no evidence for non-inflated hot Jupiters at fluxes greater than $F \sim 1.6 \times 10^{6} \si{Wm^{-2}}$ and masses of $0.37-0.98\si{\Mj}$. In this case we define non-inflated to mean that their radii could be reproduced by an evolutionary model that only considers incident flux energy deposition in the photosphere, with no additional inflationary effects \citep{fortney2007}.
\end{itemize}

To extend this study, a mixture model could be a useful tool to provide a more quantitative measure of the fraction of gas giants that do not exhibit any inflation, and could be incorporated within a Bayesian hierarchical model \citep[see][]{mixture_models_2000,celeux2007}. With increasingly detailed inflated model predictions given for wide ranges of latent parameters (such as those in \citet{fortney2007} for core mass and age), we will begin to extract information about hot Jupiter inflation from the entire observed distribution and its scatter. The forward model of our HBM, $p(R | M,F,\boldsymbol{\alpha})$, may be replaced by a theoretically-driven model for model selection purposes, and/or to constrain latent parameters in such models. Using a mixture model, we could also find the fraction of objects that are consistent with a particular model, and in which regions of mass, incident flux, and other parameters this may be the case. With the increasingly large number of discovered exoplanets, such statistics-based studies will continue to become more important.



\begin{acknowledgements}

We thank the referee, Mathieu Havel, for a comprehensive review that improved our paper. We thank Kevin Heng for valuable discussions and ideas.
M.S acknowledges support from the Swiss National Science Foundation (PP00P2-163967). B.-O.D. acknowledges support from the Swiss National Science Foundation in the form of a SNSF Professorship (PP00P2-163967). This work has been carried out within the framework of the NCCR PlanetS supported by the Swiss National Science Foundation.
Calculations were performed on UBELIX (http://www.id.unibe.ch/hpc), the HPC cluster at the University of Bern. This research has made use of the Exoplanet Orbit Database and the Exoplanet Data Explorer at exoplanets.org. This research has made use of the NASA Exoplanet Archive, which is operated by the California Institute of Technology, under contract with the National Aeronautics and Space Administration under the Exoplanet Exploration Program.\par

\end{acknowledgements}


\bibliographystyle{aa.bst}
\bibliography{exoplanets.bib} 

\end{document}